# Hyperbolic Metamaterials: From Dispersion Manipulation to Application

Zhiwei Guo, Haitao Jiang [a)], and Hong Chen [b)]

*Key Laboratory of Advanced Micro-structure Materials, MOE, School of Physics Science and Engineering, Tongji University, Shanghai 200092, China*

**OUTLINE**

I. Introduction
II. Basic physics of hyperbolic metamaterials
    A. Physical properties
        1. Dispersion relation of the anisotropic materials
        2. Enhanced spontaneous emission
        3. Abnormal refraction, reflection, and scattering
    B. Realization ways
        1. Effective medium theory
        2. Hyperbolic metasurface
        3. Natural hyperbolic media
III. Topological transition of dispersion
    A. Dispersion transition from closed ellipsoids to open hyperboloids
        1. Materials dispersion steered topological transition
        2. Loss induced topological transition
        3. Actively controlled topological transition
    B. Transition points at two kinds of topological transition
        1. Anisotropic zero-index metamaterials
        2. Linear crossing metamaterials
IV. Dispersion control in Hypercrystals
    A. Controlling the dispersion of band structure
    B. Cavity modes and edge modes with special dispersion
V. Applications
    A. Hyperlens
    B. Long-range energy transfer
    C. High sensitivity sensors
VI. Conclusion

______________________________

a) jiang-haitao@tongji.edu.cn

b) hongchen@tongji.edu.cn.

# Abstract

Manipulating the property of iso-frequency contour (IFC) will provide a powerful control for the interaction between light and matter. Importantly, hyperbolic metamaterials (HMMs), a class of artificial anisotropic materials with hyperbolic IFC have been intensively investigated. Because of the open dispersion curves, HMMs support propagating high-$k$ modes and possess enhanced photonic density of states. As a result, HMMs can be utilized to realize hyper-lens breaking the diffraction limit, meta-cavity laser with subwavelength scale, high sensitivity sensor, long-range energy transfer and so on. In order to make it easier for people who are about to enter this burgeoning and rapidly developing research field, this tutorial article not only introduces the basic physical properties of HMMs, but also discusses the dispersion manipulation of HMMs and HMM-based structures such as hypercrystals. The theoretical methods and experimental platforms are given in this tutorial. Finally, some potential applications associated with hyperbolic dispersion are also introduced.

# 1. Introduction

Recently, the hyperbolic metamaterials (HMMs) [1-4] or indefinite media [5, 6] have attracted people's considerable interest. HMM is a special kind of anisotropic metamaterial whose iso-frequency contour (IFC) is an open hyperboloid because the principal components of its electric or magnetic tensor have opposite signs [7-10]. HMMs provide some new ways to control electromagnetic waves because of their special IFC. On the one hand, by tuning the shape of hyperbolic dispersion, one can flexibly control the propagation of light in HMMs and realize the all-angle negative refraction [11-15], collimation [16-20], splitting [21-23], near-perfect absorption [24-28] and abnormal scattering [29-32]. On the other hand, in contrast to the closed IFC of normal materials (such as air), the hyperboloid IFC in HMMs is open and the large wavevectors can be supported [33-35]. As a result, the optical density of states (DOS) in the HMM can be intensively enhanced and this has important consequences for strong enhancement of spontaneous emission [36-40] and Cherenkov emissions with low energy electrons [41-44]. In addition, because the evanescent waves with large wavevectors in normal materials can become propagating waves in this strongly anisotropic medium, HMMs can realize the super-resolution imaging that overcomes the diffraction limit [22, 45-49]. Recently it also has been demonstrated that HMMs can overcome the short-range limitation of near-field coupling and realize the long-range dipole-dipole interactions [50-53]. Specially, with the aid of HMM in the dipole-quadrupole interactions, long-range electromagnetic induced transparency has been theoretically proposed and experimentally demonstrated [54].

It has been discovered that some natural materials and artificial structures can possess the hyperbolic dispersion. Owing to the excitation of the phonon polariton, some natural materials (such as graphite, SiC, $Bi_2Se_3$ and h-BN) can possess the hyperbolic dispersion in the spectrum of infrared and visible range [55-67]. Besides natural hyperbolic materials, artificial structures with subwavelength unit cells can also have hyperbolic dispersion [68-71]. Based on the effective medium theory (EMT), HMMs have been widely constructed in different spectrum by the metal-dielectric structures [72-74], the metallic nanowire structures [75-79], the multilayer fishnet [80], the circuit systems [18, 19, 22, 54, 81, 82] and the uniaxial metasurfaces [83-86]. Beyond EMT, hyperbolic dispersion can be manipulated based on the nonlocal effect of the surface waves in the metasurfaces [87-91].

Very recently, some composite structures with HMMs, including cavities and laser [92-94], hypercrystals [95-96], waveguides [97-100] and so on, have been revealed to possess unusual properties to control the electromagnetic waves. For example, in contrast to the conventional cavities, the HMM-based cavities have size-independent resonance frequencies and anomalous scaling. Therefore, they can be used to construct the subwavelength cavities. The waveguides composed of HMMs can strongly confine the guide modes and they can be used for rainbow trapping [101,102]. The metasurfaces with HMMs have been utilized for the asymmetric transmission [103] and high-efficiency wave-front manipulation [104]. The research scope of HMMs also has been extended to the actively controlled structures with graphene and $VO_2$ [105-107], in which the hyperbolic dispersion and the associated properties can be actively tuned by changing the external field. In addition, many interesting phenomena and properties have been discovered in active HMMs, including the enhanced nonlinear effect [108-110], enhanced magneto-optical effect [111-114] and topological edge states [115,116]. The above mentioned properties and applications also have been extended from photonics to acoustic [117-121] and elastic waves [122-124].

Although many systems can realize HMMs, in this tutorial we mainly discuss two kinds of systems to implement HMMs. One system to realize HMMs is microwave circuits in which the unusual physical properties of HMMs can be clearly observed. The other system is multilayered structure to realize HMMs at visible wavelengths. Many optical devices such as perfect optical absorber and high-sensitivity biosensor can be designed based on this multilayered structure.

## 2. Basic physics of hyperbolic metamaterials

### 2.1 Physical properties

**2.1.1 Dispersion relation of the anisotropic materials**

The electromagnetic response of materials depends on the permittivity and permeability [125-130]. The relative permittivity and permeability tensors of an anisotropic material are given by:

$$\hat{\varepsilon} = \begin{pmatrix} \varepsilon_\perp & & \\ & \varepsilon_\perp & \\ & & \varepsilon_{//} \end{pmatrix}, \hat{\mu} = \begin{pmatrix} \mu_\perp & & \\ & \mu_\perp & \\ & & \mu_{//} \end{pmatrix}, \quad (1)$$

where the subscripts ⊥ and // indicate the components perpendicular and parallel to the optical axis ($z$ axis), respectively. Without considering the nonlocal effect and the electro-magneto coupling, the dispersion relation of materials can be determined by substituting Eq. (1) into Maxwell's equations. For the electric HMMs ($\varepsilon_{//}\varepsilon_\perp < 0$) with isotropic permeability $\mu_{//} = \mu_\perp = 1$, the dispersion relation can be written as [1-4]:

$$(k_x^2 + k_y^2 + k_z^2 - \varepsilon_{//}k_0^2)(\frac{k_x^2 + k_y^2}{\varepsilon_\perp} + \frac{k_z^2}{\varepsilon_{//}} - k_0^2) = 0, \quad (2)$$

where $k_x$, $k_y$ and $k_z$ are the $x$, $y$ and $z$ components of the wave-vector, respectively. $k_0$ is the wave-vector in free space. Specially, the dispersion relation of the transverse-electric (TE) polarized wave ($H_x, H_y, E_z$) and transverse-magnetic (TM) polarized wave ($E_x, E_y, H_z$) are respectively described by the expressions in the first and second brackets in Eq. (2). The first kind of electric HMMs is $\varepsilon_{//} < 0, \varepsilon_\perp > 0$, where TE and TM modes can coexist. The IFC of TE mode corresponds to a sphere. However, the IFC of TM mode corresponds to a two-fold hyperboloid where two hyperbolic planes are along the direction of the optical axis and the corresponding material is called the dielectric-type or type-I HMM, as is shown in Fig. 1(a). The second kind of the electric HMMs is $\varepsilon_{//} < 0, \varepsilon_\perp > 0$, where TE mode is absent. For TM mode, the IFC has a one-fold hyperboloid and this kind of electric HMM is called the metal-type or type-II HMM, as is shown in Fig. 1(b).

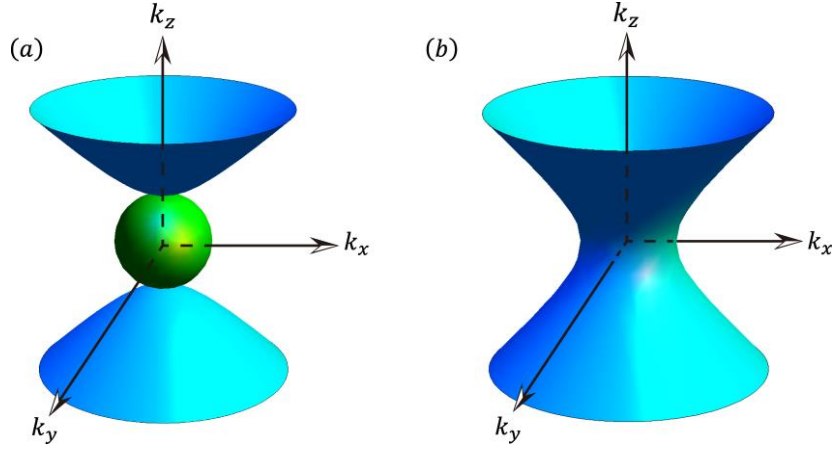

FIG. 1. IFCs of the dielectric-type (a) and metal-type (b) HMMs. The green sphere in (a) denotes the IFC of TE (TM) modes for electric (magnetic) HMM.

According to the dual principle of electromagnetic fields, for the magnetic HMMs ( $\mu_{//}\mu_{\perp}<0$ ) with isotropic permeability $\varepsilon_{//}=\varepsilon_{\perp}=\varepsilon$, the dispersion relation can be shown as:

$$(k_x^2+k_y^2+k_z^2-\mu_{//}k_0^2)(\frac{k_x^2+k_y^2}{\mu_{\perp}}+\frac{k_z^2}{\mu_{//}}-\varepsilon k_0^2)=0. \tag{3}$$

In Eq. (3), the dispersion relation of the TM and TE polarized waves are described by the expressions in the first and second brackets, respectively. Comparing Eq. (2) with Eq. (3), we can obtain similar IFCs in Fig. 1 for the magnetic HMMs. So both the dielectric-type and metal-type HMMs can be effectively controlled by tuning the signs of the permittivity and permeability.

**2.1.2 Enhanced spontaneous emission**

For normal materials with closed IFCs (such as air), the allowed modes are limited. But for HMMs, the IFCs are the open hyperboloids. As a result, the modes with large wavevectors can be supported, which leads to the enhancement of optical DOS [38, 131]. When the frequency increases from $\omega$ to $\omega+\delta\omega$, the number of allowed optical modes in the wavevector interval $\delta k_x \delta k_y \delta k_z$ is

$$\delta N=(V\delta k_x \delta k_y \delta k_z)/(2\pi)^3, \tag{4}$$

where *V* denotes the volume of the material. Take the isotropic non-magnetic materials as an example, the IFC is the closed sphere. In the infinitely thin spherical shell from $k$ to $k+\delta k$, $\delta k_x \delta k_y \delta k_z = 4\pi k^2 \delta k$. In this case, the number of modes can be derived as [131]:

$$\delta N = (V\varepsilon^{3/2}\omega^2)/2\pi^2 c^3 \delta\omega, \tag{5}$$

where $\varepsilon$ is the permittivity and $c$ is the velocity of light in vacuum. Based on Eq. (5), we can find that optical DOS in the materials with spherical IFC is a finite value because the closed IFC is integrable. Similarly, the optical DOS in the materials with a closed elliptical IFC is also a finite value. However, for the open IFC, the volume integral of the shell from $k$ to $k+\delta k$ is divergent and the optical DOS is infinite, i.e., $\delta N(k) \to \infty$. Figure 2 shows two kinds of anisotropic materials with closed IFC and open IFC, respectively. By comparing Fig. 2(a) with Fig. 2(b), we can clearly see that HMM has a diverging shell volume in the lossless limit, which means the HMM can support an infinite optical DOS. Therefore, the materials with open IFCs provide an important way of controlling the interaction between light and matter, as we will show later.

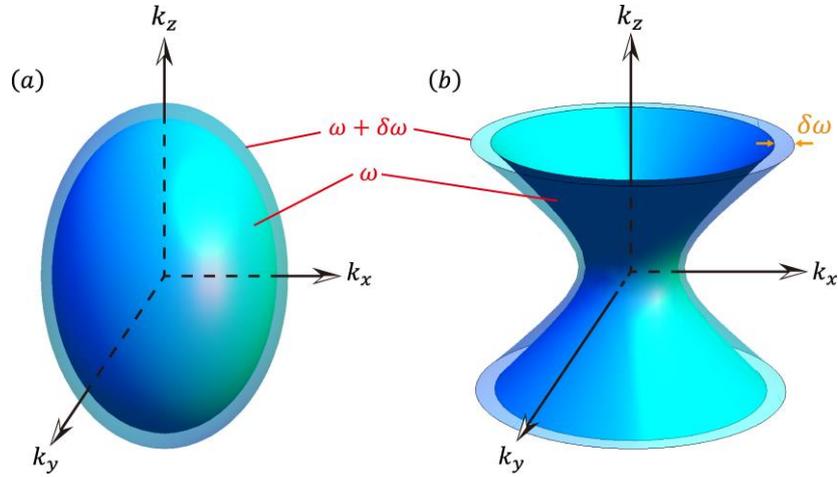

FIG. 2. The variation of IFCs for the closed ellipsoid (a) and open hyperboloid (b) when the frequency increases from $\omega$ to $\omega+\delta\omega$.

Spontaneous emission (SE), as one of the typical quantum-optic phenomenon, plays an important role in the light-emitting devices. The attenuation rate can be express as: [132-134]:

$$\Gamma_T = \frac{1}{\tau} = \frac{2\pi}{\hbar}\left|\langle f|\vec{d}\cdot\vec{E}|i\rangle\right|^2 \rho(\hbar\omega), \tag{6}$$

Where $\tau$ is the lifetime of SE and it is directly dependent on the optical DOS. $\langle f|\vec{d}\cdot\vec{E}|i\rangle$ is the transition matrix element and $\rho(\hbar\omega)$ denotes the optical DOS. The enhancement effect of SE can be represent by the Purcell factor $F_p = \Gamma_T/\Gamma_0$, where $\Gamma_0$ is the attenuation rate of vacuum [135]. From Eq. (6), we can clearly see that $F_p$ strongly depends on the optical DOS of the surrounding medium. The optical DOS $\rho(\hbar\omega)$ is equal to $\sum_k \hbar\delta(\omega_k - \omega)$ and it can be effectively tuned by changing the shell volume enclosed by the corresponding IFCs [1]. So, along with the topological transition of materials from closed IFCs to the open IFCs, the SE in medium will be significantly enhanced because optical DOS will gradually diverge.

In 2011, researchers theoretically studied the Purcell factor of an electric dipole such as quantum dot with a finite volume in HMMs [136]. The radiative lifetime is:

$$\frac{1}{\tau_\alpha} = -\frac{8\pi d^2 q_0^2}{\hbar} \int \frac{d^3k}{(2\pi)^3} \operatorname{Im} G_{k,\alpha\alpha} \Phi_k^2, \alpha = x, y \text{ or } z, \qquad (7)$$

where $q_0 = \omega/c$ and $d$ are the wavevector in vacuum and the effective matrix element of the dipole moment, respectively. $\Phi_k$ denotes the distribution of the emitter polarization. $G_k$ is the Green function in the $k$ space. The lifetime $\tau_x = \tau_y$ and $\tau_z$ describe the decay of the source initially polarized in the plane $x-y$ and along the $z$ axis, respectively [136]. The larger the Purcell factor is, the more the high-level photon states will transit to the low-level in a period of time. Therefore, the intensity of the radiated light will decay faster and the radiative lifetime will be shorter. Reducing the radiation lifetime is of great significance for improving fluorescence efficiency [137-140]. The permittivity and permeability of HMM are assumed to be [136]

$$\varepsilon_\perp(\omega) = \varepsilon_\perp^0 + \frac{\Gamma_0}{\omega_0 - \omega - i\Gamma}, \varepsilon_{//} = 1, \mu = 1. \qquad (8)$$

The dependence of $\varepsilon_\perp$ on the normalized frequency $(\omega_0 - \omega)/\Gamma_0$ is shown in Fig. 3(a). The red, orange, yellow, green, cyan, and blue lines correspond to $\Gamma/\Gamma_0 = 0.01, 0.015, 0.02, 0.03, 0.05$ and 0.1, respectively. The real part and imaginary part of $\varepsilon_\perp$ are marked by solid and dashed lines, respectively. With the increase in the value of $(\omega_0 - \omega)/\Gamma_0$, the sign of $\varepsilon_\perp$ will change. When the sign of $\varepsilon_\perp$ and $\varepsilon_{//}$ are same, the IFC of system corresponds to the elliptical dispersion. However, when the sign of $\varepsilon_\perp$ and $\varepsilon_{//}$ are opposite the IFC of system corresponds to the hyperbolic dispersion. The Purcell factor for the different dipole orientations in material as a function of $(\omega_0 - \omega)/\Gamma_0$ and $\Gamma/\Gamma_0$ are shown in Figs. 3(b) and 3(c), respectively. The dipole orientations along $z$ and $x$ are marked by cyan arrow and pink arrow, respectively. It is seen that Purcell factor enhances greatly in hyperbolic region than elliptical region. In order to see this feature more clearly, $\Gamma/\Gamma_0$ is fixed at 0.03. When the direction of dipole along the $z$ direction, the dependence of the Purcell factor on the normalized frequency are shown in the inset of Fig. 3(b). Similar to Fig. 3(b), when the direction of dipole changes to $x$, the Purcell factor also will be greatly enhanced in the hyperbolic regime, as shown in Fig. 3(c). The increased rates of spontaneous emission of emitters have been experimentally observed [38]. The CdSe/ZnS colloidal quantum dots are placed in the near field of the HMM, which is realized by the multilayered structure composed of silver and titanium dioxide (TiO$_2$) thin films, as is schematically shown in Fig. 3(d). Due to the dispersion of the silver, the effective parameters is frequency dependent. In Fig. 3(e), with the increase of wavelength, multilayered structure will undergo a topological transition from the closed IFC to open IFC [38]. This topological transition of the IFC leads to a significant decrease in the spontaneous emission lifetime of quantum dots, as depicted in Fig. 3(f). The reason for lifetime decrease is the increased optical DOS when the system goes through the topological transition of IFC from a closed ellipsoid to an open hyperboloid [38].

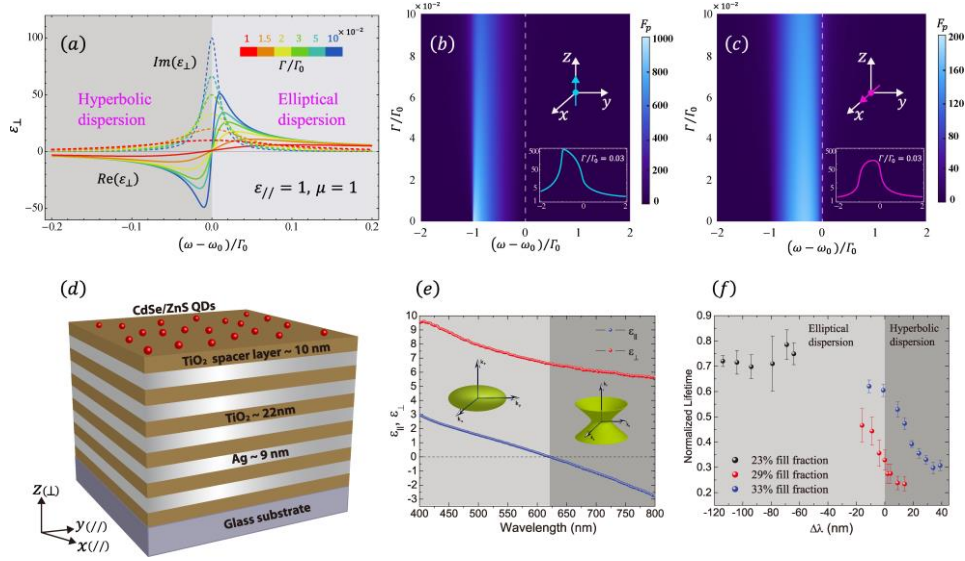

FIG. 3. (a) The perpendicular component of permittivity as a function of the normalized frequency. The solid and dashed lines correspond to the real and imaginary parts, respectively. (b) The Purcell factor as a function of the normalized frequency and loss when a dipole source along $z$ direction is placed in the material. (c) Similar to (b), but the direction of dipole source changes to $x$. Reproduced with permission from Phys. Rev. A 84, 023807 (2011). Copyright 2011, American Physical Society. (d) Schematic of the quantum dots on the surface of material. (e) Effective perpendicular and parallel components of permittivity for the multilayered structure. (f) Radiative lifetime in elliptical and hyperbolic dispersion regions. Reproduced with permission from Science 336, 205 (2012). Copyright 2012, American Association for the Advancement of Science.

The radiation patterns of point source in materials can visually represent the IFC of materials. Based on the theory of Green's function, the dipole emission in materials can be obtained [141]. Specially, the emission patterns of HMMs have been theoretically calculated [142]. The two dimensional (2D) IFC, in which we only need study the relationship between $k_x$ and $k_z$, is used to analyze the light emission. The 2D IFC of isotropic material ($\varepsilon=1, \mu=1$) is shown in Fig. 4(a), where the group velocity (density of energy flow) are marked by the green arrows. By putting a dipole source into the structure, the corresponding calculated radiation patterns are presented in Fig. 4(e). Three different anisotropic materials with elliptical ($\varepsilon=1, \mu_x=1, \mu_z=2$), hyperbolic ($\varepsilon=1, \mu_x=-2, \mu_z=1$) and linear-crossing ($\varepsilon \to 0, \mu_x=-2, \mu_z=1$) dispersions are shown in Figs. 4(b)-4(d), respectively. By comparing Fig. 4(d) with Fig. 4(c), one can see that the topological transition of IFC will happen once $\varepsilon>0, \mu_x\mu_z<0$. Moreover, the IFC will become two linear-crossing lines when the parameters change to $\varepsilon \to 0, \mu_x\mu_z < 0$ [22]. In Figs. 4(f)-4(g), the corresponding calculated emission patterns for different parameters coincide well with the IFCs in Figs. 4(b)-4(d), respectively. For example, for the case with elliptical dispersion (Fig. 4(b)), the light can propagate in all in-plane directions (Fig. 4(f)). However, for the case with open hyperbolic dispersion ((Fig. 4(c))), the light can only propagate in a certain range of angles (Fig. 4(g)) and the field vertical to the hyperbolic asymptotes is much stronger than in other places because of the larger optical DOS. In particular, for the case with linear-crossing dispersion, the light is collimated and only propagate in two directions [22]. In experiment, the emission patterns can be measured when a point source put into the inner of the structures [19, 81, 82, 143, 144]. So the light emission patterns can be effectively controlled by the IFCs of the materials. Importantly, the measured field distributions can be used to determine the IFC of materials by using the Fourier transform [145, 146].

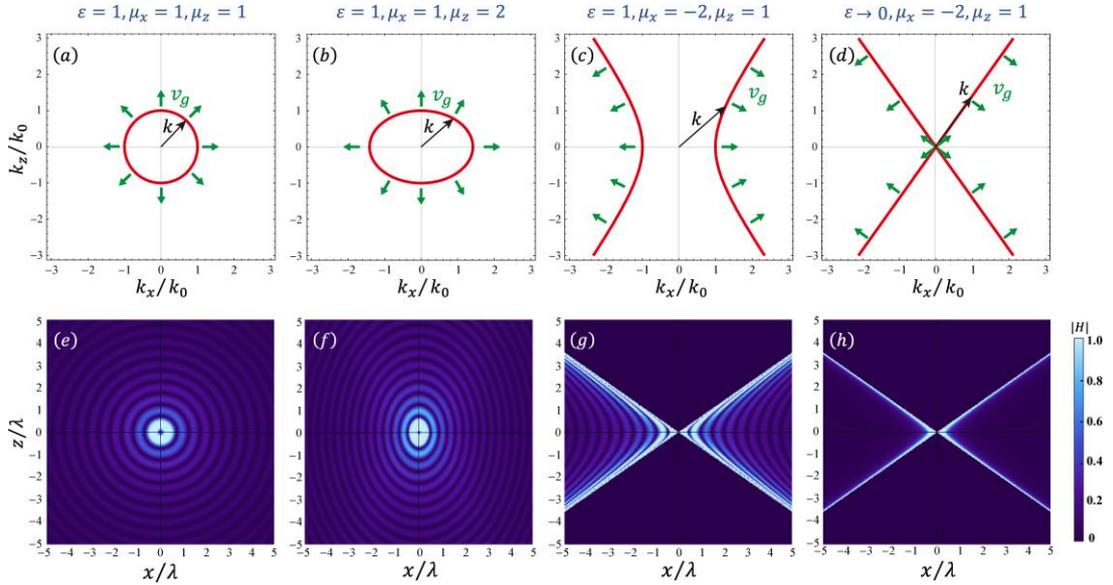

FIG. 4. IFCs (the upper row) and the corresponding in plane magnetic field $|H|$ distributions (the lower row). The green arrows in upper row denote group velocity. (a) $\mu_x = \mu_z$, corresponds to the isotropic material. (b-e) $\mu_x \neq \mu_z$, corresponds to the elliptical, hyperbolic and linear-crossing

anisotropic materials, respectively.

**2.1.3 Abnormal refraction, reflection, and scattering**

Metamaterials, artificial materials composed of subwavelength unit cells, provide a powerful platform to manipulate the propagation of light [125-130]. The counterintuitive negative refraction and the superlens beyond the diffraction limit were initially confirmed to exist in left-handed metamaterials [125, 126]. After that many charming optical phenomena have been discovered, such as the cloaking using transform optics [127], optical tunneling [147], and trapping light [148-150]. In recent years, HMMs have been proved to have unprecedented abilities to tune the electromagnetic waves and can realize many unusual optical phenomena such as negative refraction [11-15], near-perfect absorption [24-28], abnormal scattering [29-32] and directional propagation [22]. Unlike the left-handed metamaterials and the metamaterials with complex electromagnetic distribution, HMMs is very easy to realize as long as the signs of the EM parameters principal components are opposite [1-4]. Either positive or negative refractions of light can exist in the HMMs and the IFCs of materials can be used to analyze the abnormal optical properties.

To illustrate this point clearly, four kinds of magnetic HMMs are shown in Fig. 5. All four HMMs have the normal dispersion characteristics $\varepsilon(\omega+\delta\omega) > \varepsilon(\omega), \mu(\omega+\delta\omega) > \mu(\omega)$, that means the permittivity and permeability will increase with the increase of frequency. $\delta\omega$ denotes a tiny increase in frequency. The IFCs of four HMMs are shown in Figs. 5(a)-5(d), in which the IFCs of the HMM (air) at frequency $\omega$ and $\omega+\delta\omega$ are respectively marked by the solid blue (red) and dashed blue (red) lines. The light is incident from the air (the lower part) to the HMMs (the upper part) at an oblique angle of 20 degrees, which are marked by the red arrow of wavevector $k_1$. Based on the law of conservation of tangential wavevector and law of causality, the wavevector $k_2$ and Poynting vector $S$ in HMMs can be determined, as is shown in Figs. 5(a)-5(d). For the first case (HMM1: $\varepsilon_y > 0, \mu_x > 0, \mu_z < 0$), it belongs to the dielectric-type HMMs, in which negative refraction phenomenon will occur. However, unlike the negative refraction of left-handed materials [151, 125], the directions of wavevector and energy flow in HMMs form a certain angle rather than a complete antiparallel, which is shown in Fig. 5(e). For the second case (HMM2: $\varepsilon_y < 0, \mu_x < 0, \mu_z > 0$), it also belongs to the dielectric-type HMMs. Under this condition, the dashed line is outside the solid line in Fig. 5(b), which means the IFC expands with the increase of frequency. In this case, the light incident on this HMM will undergo positive refraction, as shown in Fig. 5(f). For the third (HMM3: $\varepsilon_y > 0, \mu_x < 0, \mu_z > 0$) and fourth (HMM4: $\varepsilon_y < 0, \mu_x > 0, \mu_z < 0$) cases, they belong to the metal-type HMMs. Positive and negative refraction can respectively occur in HMM3 and HMM4 only when the tangential wavevector in the air is greater than a critical value. Otherwise, when the incident angle is less than the critical angle the incident wave will be totally reflected. In particular, in Figs. 5(g) and 5(h) the Goos-Hänchen shift associated with total reflection in HMM3 and HMM4 are positive and negative, respectively. Recently, the refractive index sensing is demonstrated by exploiting the singular phase of light associated with the enhanced Goos-Hänchen shift in HMMs [152-155].

In addition to the anomalous refraction and reflection, HMMs have also been shown to manipulate some novel scattering behaviors of electromagnetic waves [30-32]. For example the anomalously weak scattering can be achieved when the impedance of the multilayer HMMs matches with the surrounding medium [30]. The multifrequency superscattering (by overlapping multiple decoupled resonances in subwavelength object) beyond the single-channel limit [156] have been proposed based on the subwavelength hyperbolic structures [31]. In particular, for lossy HMMs, scattering caused by radiation delay channels and absorption caused by non-radiation delay channels can be effectively controlled [28, 137].

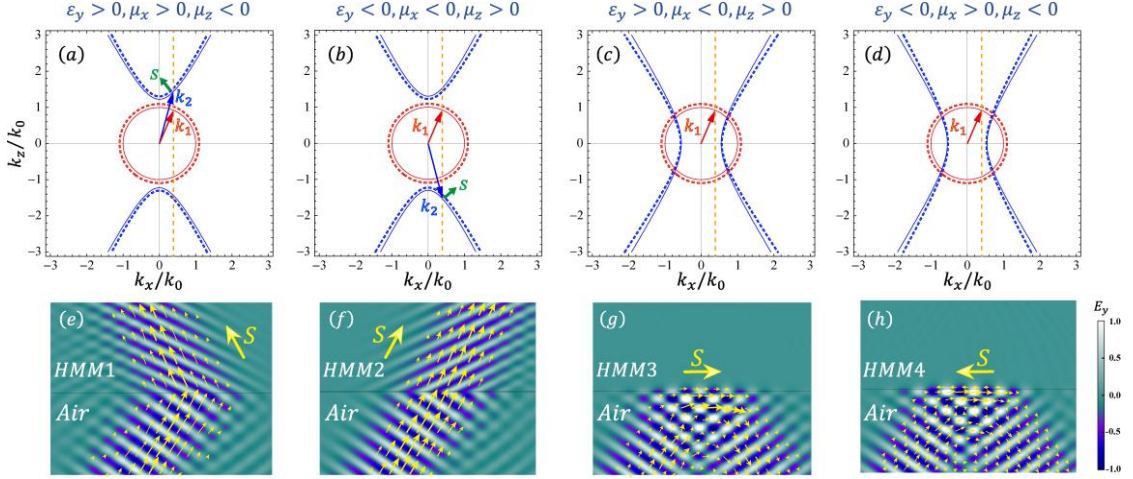

FIG. 5. IFCs (the upper row) and the corresponding electric field $H_y$ distributions (the lower row). The red and blue arrows in upper row denote the wavevectors in air and HMMs, respectively. The green arrows in upper row denote group velocity in HMMs. In (e)-(h), the light incident from the air to the HMMs and the yellow arrow denote the group velocity in HMMs.

## 2.2 Realization ways

### 2.2.1. Effective medium theory

Based on EMT, the periodic arrangement of artificial structures with sub-wavelength unit-cell size can be regarded as an effective homogeneous medium, which can be characterized by macroscopic homogeneous permittivity and permeability [5, 6, 157]. By designing suitable artificial structures such as the metal/dielectric multilayers (Fig. 6(a)), the metal nanowire arrays (Fig. 6(b)), the multilayer fishnet (Fig. 6(c)) and the microwave circuit system loaded with lumped elements (Figs. 6(d) and 6(e)), HMMs can be conveniently realized [1-4]. Recently, the effectiveness of EMT for constructing HMMs has also been studied in depth [158-160]. EMT is a quasi-static approximation. This theory may be invalid when the optical nonlocal effect exists in the system [161]. For example, this approximation may be destroyed when surface plasmons (SPPs) are excited or when the unit-cell size is not in the sub-wavelength scale.

In the visible range, both metal/dielectric multilayers and metal nanowire arrays are used to realize the electric HMMs with anisotropic permittivity according to EMT. Specifically, the effective parameters of metal/dielectric multilayers structure in Fig. 6(a) can be expressed by [72]:

$$\varepsilon_{//} = p\varepsilon_m + (1-p)\varepsilon_d, \varepsilon_{\perp} = \frac{\varepsilon_m \varepsilon_d}{p\varepsilon_d + (1-p)\varepsilon_m}, \quad (9)$$

where $p = t_m/(t_m + t_d)$ means the filling ratio of metal layer, $\varepsilon_m$ ($t_m$) and $\varepsilon_d$ ($t_d$) are the permittivity (thickness) of metal layer and dielectric layer, respectively. Once the condition $\varepsilon_{//}\varepsilon_{\perp} < 0$ is satisfied by tuning the parameters, the metal/dielectric multilayer structure can be seen as an electric HMM. The metal nanowire arrays in Fig. 6(b) also can be used to construct the electric HMMs, in which the effective permittivity are given by [162]:

$$\varepsilon_{//} = \frac{[(1+p)\varepsilon_m + (1-p)\varepsilon_d]\varepsilon_d}{(1-p)\varepsilon_m + (1+p)\varepsilon_d}, \varepsilon_{\perp} = p\varepsilon_m + (1-p)\varepsilon_d, \quad (10)$$

where the filling ratio of metal wires in nanowire arrays is $p = A/A_0$. $A$ and $A_0$ are the cross-sectional area of metal wires and host material, respectively. Similarly, the sign of effective $\varepsilon_{//}$ and $\varepsilon_{\perp}$ can be tuned by changing the structure parameters.

Besides the electric HMMs, recently the magnetic HMMs with anisotropic permeability have also attracted people's great attention [7-10, 22, 54]. For the multilayer fishnet metamaterial composed of metal and dielectric layers in Fig. 6(c), both the permittivity and permeability can be tuned by changing the structure parameters or choosing different wavelength [7]. In Ref [7], the real part of $\varepsilon_{//}$ and $\varepsilon_{\perp}$ are positive and negative at 1310 nm. In this case, the condition of $\varepsilon_{//}\varepsilon_{\perp} < 0$ is satisfied and multilayer fishnet corresponds to an electric HMM. However, when the wavelength is 1530 nm, the condition of $\mu_{//}\mu_{\perp} < 0$ is satisfied and a magnetic HMM is realized [7]. In a different way, magnetic HMMs can be easily realized by circuit-based circuit metamaterials in microwave regime [22, 54]. The circuit system can support quasi-TEM wave when the thickness of dielectric plate is much less than the wavelength. Based on the circuit-based system, the relationship of electric and magnetic fields can be easily mapped by the relationship between voltage and current in the circuit [163, 164]. As a result, the electromagnetic response is equivalent to the circuit parameters. In Fig. 6 (d), the impedance and admittance of the circuit are represented by $Z$ and $Y$, respectively. The direction of the magnetic field produced by the current can be determined according to Ampere's law, as is shown in Fig. 6(d). By mapping the circuit equation (telegraph equation) to the Maxwell equation, the relationship between circuit and electromagnetic parameters are described by :

$$Z = i\omega\mu, \ Y = i\omega\varepsilon. \tag{11}$$

For the circuit without lumped elements, $Z = i\omega L_R$ and $Y = i\omega C_R$. When the microwave circuits loaded with series-lumped capacitors $C_L$ and shunted-lumped inductors $L_L$:

$$Z = i\omega L_R + \frac{1}{i\omega C_L}, \ Y = i\omega C_R + \frac{1}{i\omega L_L}. \tag{12}$$

So the effective electromagnetic parameters of circuit with lumped elements take the form:

$$\mu = i\omega L_R - \frac{1}{\omega^2 C_L}, \ \varepsilon = C_R - \frac{1}{\omega^2 L_L}. \tag{13}$$

From Eq. (13), we can clearly see that the effective permittivity and permeability of circuit system can be easily tuned by the lumped elements in the circuit. Figure 6(e) shows a simple effective circuit model for the circuit-based HMM. In this circuit model, capacitors are loaded in the $x$ direction to realize the anisotropy of permeability and $\mu_x = L_R / g\varepsilon_0$. In the long-wavelength limits, if we do not consider the loss, the effective permittivity and permeability of 2D circuits can be written as [22]:

$$\begin{aligned} \varepsilon &= (2C_R g - g / \omega^2 L_L d) / \varepsilon_0, \\ \mu_y &= (L_R / g - 1 / \omega^2 C_L d) / \varepsilon_0, \end{aligned} \tag{14}$$

where $g$ is the structure factor that can be tuned by the thickness of dielectric layer and the width of the metal strip. $d$ is the length of a unit cell. It can be seen from Eq. (14) that $\mu_y$ and $\varepsilon$ can be independently adjusted by changing $L_L$ and $C_L$. By tuning the sign of $\mu_y$ from positive to negative value, the topological transition of IFC from closed elliptic to open hyperbolic dispersion can be realized.

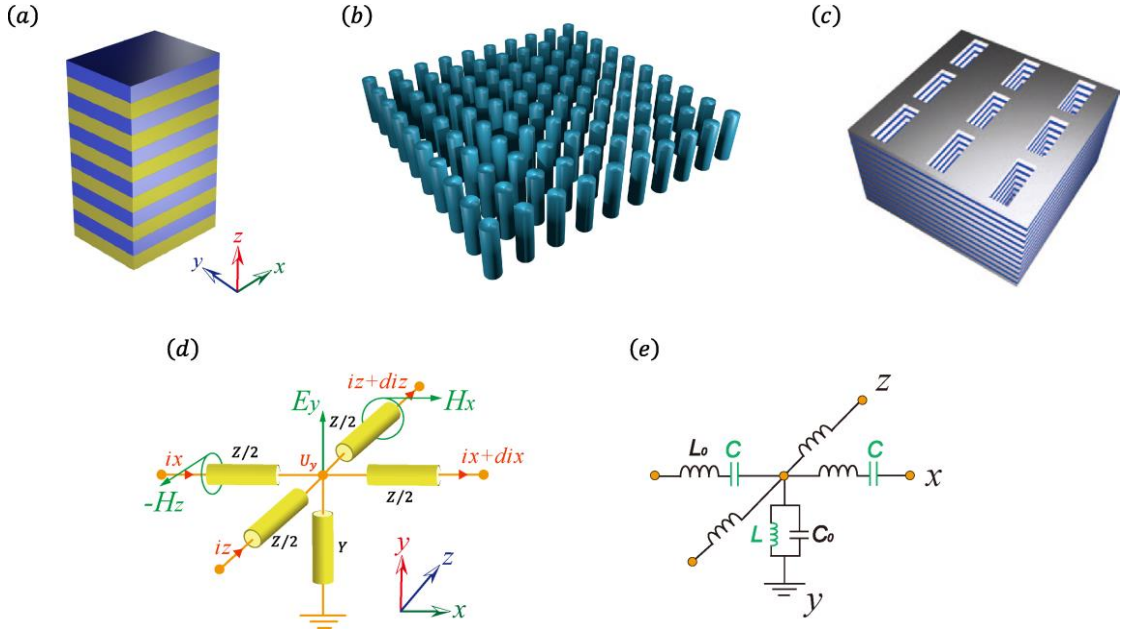

FIG. 6. Schematics of metal/dielectric multilayer HMM (a), metal nanowire HMM (b) and multilayer fishnet HMM (c) in the frequency range from THz to UV. Reproduced with permission from Nat. Photon. 7, 948 (2013). Copyright 2013, Springer Nature. (d), (e) The circuit-based HMM in microwave regime.

**2.2.2. Hyperbolic metasurface**

Flatland optics with hyperbolic metasurface (HMS) is a hot research topic at present [85, 165, 166], in which the hyperbolic dispersion can be manipulated by the in-plane SPPs. By choosing different frequencies of the surface wave, the elliptical or hyperbolic IFCs can be obtained. So far, a variety of HMSs have been proposed from visible light to microwave regime [165-171]. One of important HMSs is the metal nano-grating array, which can effectively manipulate the SPP in the near-field region and the hyperbolic dispersion can be realized in this structure [14, 167]. The schematic of the metal/air grating is shown in Fig. 7(a). The corresponding IFCs at different frequencies are given in Fig. 7(b). Based on the IFCs, the positive refraction, collimation, and negative refraction for the red, green, and blue light can be determined [14]. Moreover, by using the physical mechanism of spin-orbit coupling of evanescent waves [172-175], the photonic spin Hall effect (the photons with different circular polarizations will go different ways) have been experimentally observed [14]. In addition to the metal nano-grating array, HMS also can be realized by the anisotropic conductivity layer, as is shown in Fig. 7(d), in which a conductivity layer is sandwiched by two medium with permittivity $\varepsilon_1$ and $\varepsilon_2$, respectively [87]. The dispersion conductivity at different directions in plane $yoz$ are shown in Fig. 7(e). With the change of frequency, the imaginary parts of $\sigma_{//}$ and $\sigma_\perp$ can be effectively tuned. The cases $\text{Im}(\sigma_\perp) > 0, \text{Im}(\sigma_{//}) > 0$ and $\text{Im}(\sigma_\perp) < 0, \text{Im}(\sigma_{//}) < 0$ correspond to the inductive (as a metal sheet) and capacitive (as a dielectric sheet) metasurface [87]. So, when both the $\text{Im}(\sigma_\perp)$ and $\text{Im}(\sigma_{//})$ are negative, the IFCs of the surface plasmon is an elliptic and the excited plasmon can propagate in all the directions of $yoz$ plane. However, when one of the $\text{Im}(\sigma_\perp)$ and $\text{Im}(\sigma_{//})$ is positive, the IFCs of the surface plasmon are hyperbolas. Moreover, two kinds of hyperbolic dispersion with different principal axis can be realized when $\text{Im}(\sigma_\perp) > 0, \text{Im}(\sigma_{//}) < 0$ or $\text{Im}(\sigma_\perp) < 0, \text{Im}(\sigma_{//}) > 0$. As a result, the emission patterns of a dipole in this HMS can be controlled, as is shown in Fig. 7(f). In addition, the HMS based on the conductivity tensor has also been proposed in the array of graphene strips [84-86]. In this actively controlled structure, the topological phase transition of IFC from a closed ellipse to an open hyperbola can be achieved by adjusting the chemical potential of graphene (the details about dispersion tuned by external field will be introduced in detail in section 3.1.3).

In microwave regime, by manipulating the magnetic SPP at subwavelength scale, the magnetic HMS can be easily realized [10]. In this HMSs, the arrays of coiling copper wires are fabricated on a dielectric substrate. As is shown in Fig. 7(g), the double split ring resonator is the basic unit cell. By calculating the engine-modes of the structure, the 3D dispersion relationship can be obtained. Moreover, the IFCs at a fixed frequency can be determined. In Fig 7(h), as the frequency increases, the topological transition of IFC will occur. In microwave regime, not only can the size and shape of the meta-atoms be flexibly fabricated, but also the electric and magnetic fields can be delicately measured [10]. Based on the near-field detection technology and Fourier transform, the hyperbolic IFC of HMS can be clearly observed in Fig. 7(f).

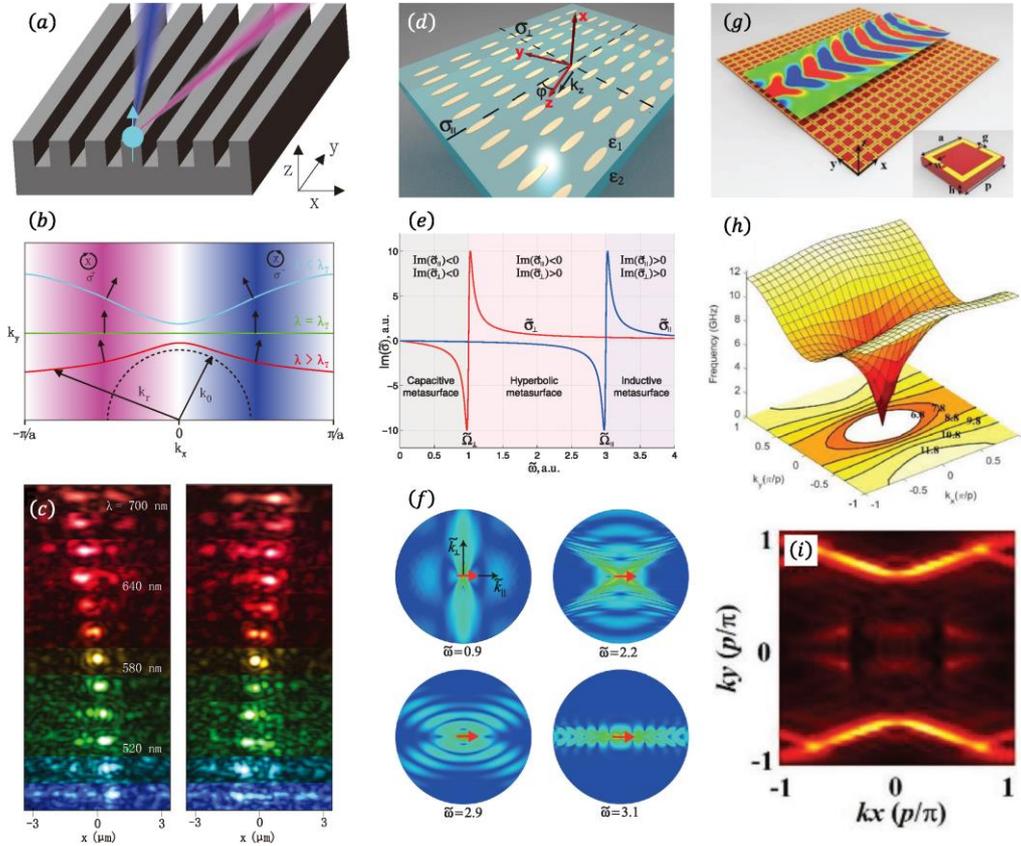

FIG. 7. HMS realized by the metal/air grating (a)-(c), the anisotropic conducting layer (d)-(f), and the array composed of split-ring-resonators (g)-(i). (b) The IFCs of (a) at different wavelength. (c) The photonic spin Hall effect in (a). Reproduced with permission from Nature 522, 192 (2015). Copyright 2015, Springer Nature. (e) The imaginary parts of anisotropic conductivity layer in (b). (f) The electric field distributions in (b) at different frequencies. Reproduced with permission from Phys. Rev. B 91, 235423 (2015). Copyright 2015, American Physical Society. (h) The 3D dispersion relationship and the 2D IFC in (g). (i) Experimentally measured IFC of (g). Reproduced under the terms of a Creative Commons AttributionNonCommercial-NoDerivs 4.0 International License. Adv. Sci. 5, 1801495 (2018). Copyright 2018, The Authors.

**2.2.3. Natural hyperbolic media**

The greatest advantage of artificial HMMs is that structural parameters can be easily and flexibly controlled. However, in high-frequency regime, because of the complex nanofabrication process, artificial HMMs have to suffer from the structure tolerances associated increased losses and the size-dependent limitations. The natural hyperbolic materials are the attractive alternative, such as the graphite, SiC, $Bi_2Se_3$, $MgB_2$ and hexagonal boron nitride (h-BN) [65, 67]. Many natural hyperbolic materials belong to the class of van der Waals (vdW) crystals (consist of individual atomic planes bonded by weak vdW forces), and the polaritons in natural hyperbolic materials possess the unprecedented control for the light-matter interaction [177, 178]. As a representative low-loss vdW materials, the h-BN supports both Type I and Type II hyperbolic phonon-polaritons at two separated Reststrahlen bands in mid-infrared regime [63, 179]. The high confined phonon-polariton in h-BN has the properties analogous to the SPPs. However, the ultra-short wavelength, and low loss compared with metal-based SPP and graphene plasmons, which makes it an excellent candidate for nano-photonics. The anisotropic permittivity of h-BN is shown in Fig. 8(a). When the signs of out of plane relative permittivity ($\varepsilon_\perp$) and in plane relative permittivity ($\varepsilon_{//}$) are different, the polaritons possess an in-plane hyperbolic dispersion, in which the iso-frequency contours are open hyberboloids. In the lower Reststrahlen band (760–825 cm$^{-1}$), $\varepsilon_\perp < 0$, $\varepsilon_{//} > 0$, and the phonon polaritons show the Type I hyperbolic dispersion. However, in the upper Reststrahlen band (1360–1610 cm$^{-1}$), $\varepsilon_\perp > 0$, $\varepsilon_{//} < 0$, and the phonon polaritons show the Type II hyperbolic dispersion [179]. Besides, special attention has been given to the long-lifetime, ultra-slow propagating, and sub-diffraction imaging in h-BN [56, 180-182]. The dispersion relation of the polariton modes in h-BN is visualized by the false-color plot of the imaginary part of the reflectivity $\text{Im} \, r_p(\omega - q)$, where $q$ denotes the momentum. Figure 8(b) shows the dispersion relation of the hyperbolic photon-polaritons modes in h-BN. The finite-thickness slab of h-BN can act as multimode waveguides for the propagation of hyperbolic phonon-polaritons. Because of the negative dispersion in the lower Reststrahlen band, the phonon-polaritons of layered h-BN slab can realize negative refraction [56]. The IFC of the hyperbolic phonon-polaritons can be controlled by changing the frequency, and the propagation angle within the two spectral bands is varying. The h-BN can be used to realize the sub-wavelength imaging in two separate spectral regions. The imaging system is composed of the Si substrate, the Au disk and the h-BN layer, as is shown in Fig. 8(c). The polaritons can be launched by the Au disk edges and propagate towards the h-BN. The electric distribution of hyperbolic phonon-polaritons at different frequencies are shown in Fig. 8(c). It is clearly show that the propagation angle of the hyperbolic polaritons depends on the frequency and it can be quantitatively inferred from the reconstructed outline $\tan \theta(\omega) = [D(\omega) - d]/2h$ [180]. Based on the polaritons, the topological transition of IFC from a closed ellipse to an open hyperbola can be realized by patterning a gating structure, where the h-BN ribbons are separated by air gaps [183]. In addition, the anomalous internal reflections in the nanocone array composed of h-BN have been experimentally observed [184]. Moreover, the design idea that pattering new structure with h-BN can be extended to other structures vdW materials (such as $MoS_2$, $Bi_2Se_3$)to manipulate the hyperbolic phonon-polaritons.

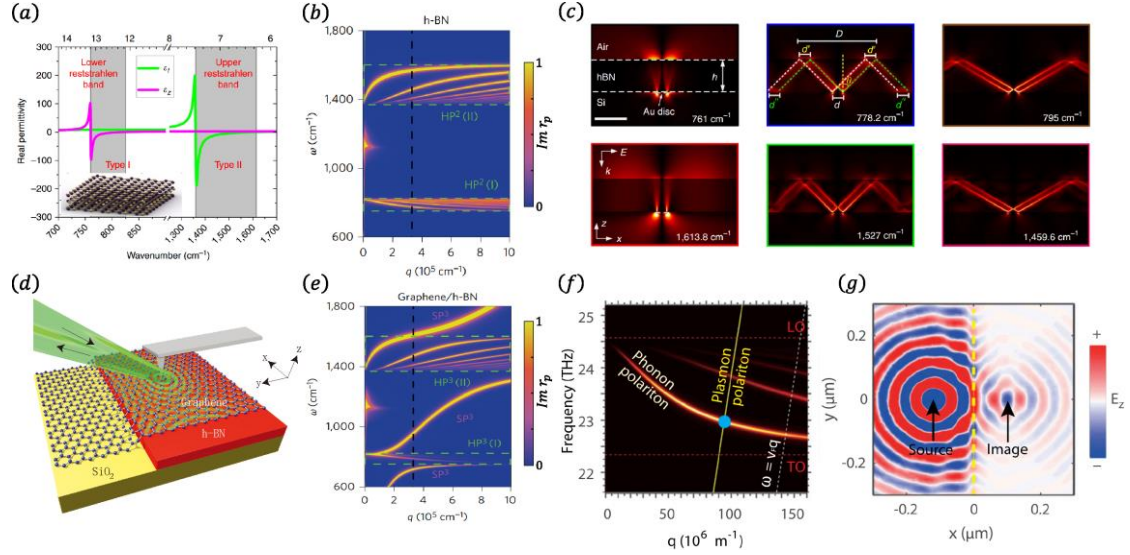

FIG. 8. The h-BN and graphene/h-BN hyperbolic configuration. (a) The real parts of permittivity tensor of h-BN. Two shadow regions are the Reststrahlen band where $\varepsilon_{\perp}\varepsilon_{//} < 0$. Reproduced with permission from Nat. Commun. 5, 5221 (2014). Copyright 2014, Springer Nature. (b) The dispersion relation of the hyperbolic phonon polaritons in h-BN of 58 nm. (d) Schematic of the scattering-type scanning near-field optical microscope in graphene/h-BN heterostructure, where the probe tip can be excited by a laser source. The scattered plasmons also are collected by the tip. (e) Dispersion relation of the phonon polaritons in a graphene/h-BN heterostructure. Reproduced with permission from Nat. Nanotech. 10, 682 (2015). Copyright 2015, Springer Nature. (c) The electric filed distribution in the h-BN. The directional angles of the hyperbolic polaritons can be tuned by choosing the various frequencies. Reproduced under the terms of a Creative Commons AttributionNonCommercial-NoDerivs 4.0 International License. Nat. Commun. 6, 7507 (2015). Copyright 2015, The Authors. (f) Dispersion relation of the graphene/h-BN heterostructure with strong coupling between the plasmon and phonon polaritons. (g) Imaging based on the all-angle refraction. In the electric field distribution, the positions of the source and image are marked. Reproduced with permission from Proc. Natl. Acad. Sci. U.S.A 114, 6717 (2017). Copyright 2017, National Academy of Sciences.

Although two separated bands of h-BN can realize the hyperbolic phonon-polaritons, the electrodynamic properties of h-BN is limited by the intrinsic crystal lattice. To overcome this limitation, the heterostructures composed of graphene and h-BN are proposed because it may provide a new regime where the SPPs(coherent oscillations of the electron density) in graphene [185] and the hyperbolic phonon polaritons (atomic vibrations) in thin h-BN are combined [60, 106, 186-188]. The schematic of graphene/h-BN heterostructure is presented in Fig. 8(d), where the hybridized plasmon–phonon modes can effective modulate the dispersion relation of SPPs and phonon polaritons [106, 189]. To explore the polaritonic response of graphene/h-BN heterostructure, polariton interferometry (scattering-type scanning near-field optical microscopy) are performed in the experiment. The hyperbolic plasnon-phonon polaritons possess the combined virtues of graphene and h-BN and its propagation length 1.5-2.0 times greater than that of hyperbolic phonon polaritons in h-BN [106]. Figure 8(e) shows the dispersion relation of the graphene/h-BN heterostructure in Fig. 8(d). By comparing Fig. 8(e) with Fig. 8(b), we can clearly see that the new collective modes, hyperbolic plasmon-phonon polaritons, can be controlled by the hybridization of SPPs and phonon polaritons. Importantly, the sign of the group velocity of the new hybrid modes also can be modulated based on the strong modes coupling [60]. In Fig. 8(e), we can clearly see that the group velocity of graphene palsmon polaritons and h-BN phonon polaritons are respectively positive and negative in the BN's first reststrahlen band. At a working frequency (22.96 THz), which corresponds to the blue point in Fig. 8(f), the wavevectors of these polaritons are equal. Figure 8(g) present the imaging of a point source in the graphene/h-BN heterostructure. The plasmons in graphene are excited by a dipole source in the left region (marked by the black arrow) and then they couple to the polaritons in h-BN. Because the group velocity in the h-BN is negative, an image of the source will occur in the right region associated with the all angle negative refraction at the interface (marked by the yellow dashed line). The highly confined low-loss plasmons in graphene/h-BN heterostructures have been experimentally confirmed [187].

Very recently, semiconducting vdW α-$MoO_3$ as a new member to the growing list of polaritons in vdW materials [61, 190, 191]. In mid-infrared regime, the thin flake α-$MoO_3$ can support the hyperbolic phonon polaritons in different Reststrahlen bands. The α-$MoO_3$ lattice is composed of octahedron unit cells with nonequivalent Mo-O bonds along the three principal crystalline axes, as is shown in Fig. 9(a) [190]. The inset graphic in Fig. 9(b) shows the unit cell of α-$MoO_3$, in which three different oxygen sites are indicated by O1–O3, respectively. In the mid-infrared region, the optical response of the α- MoO3 is dominated by the phonon absorption rather than electronic transition, the permittivity can be described by the Lorentz equation [191]. The permittivity tensor of this low-symmetry crystalline can be written as $diag[\varepsilon_x, \varepsilon_y, \varepsilon_z]$ with $\varepsilon_x \neq \varepsilon_y \neq \varepsilon_z$ and at least one of the tensors is negative, which is presented in Fig. 9(b). Three Reststrahlen bands are marked by the pink, yellow and blue shadows, respectively. It is clearly see that the sign of the $\varepsilon_i$ $(i = x, y, z)$ can be tuned by choosing different frequency. The dispersion relation of α-MoO3 can be calculated by the general Fresnel equation. From the special permittivity tensor, it can be expected that α-MoO3 is a kinds of biaxial hyperbolic crystal that four conical singularities in the $k$-space are connected, as illustrated in Fig. 9(c). In band 1 and band 2, one of the in-plane permittivity $\varepsilon_i$ $(i = x, y)$ is negative while the out-of-plane permittivity $\varepsilon_z$ is positive, which will induce the hyperbolic dispersion and elliptical dispersion in-plane and out-of-plane, respectively. By exchanging the signs of $\varepsilon_x$ and $\varepsilon_y$, the principal axis direction of hyperbolic dispersion can be changed accordingly. The hyperbolic phonon polaritons is similar to the uniaxial crystalline, such as h-BN. However, in band 3, two in-plane permittivity $\varepsilon_i$ $(i = x, y)$ are positive while the out-of-plane permittivity $\varepsilon_z$ is negative. In this case, the elliptical dispersion occur in-plane while the hyperbolic dispersion occur out-of-plane. So, the α-MoO3 provides a way to simultaneously control hyperbolic phonon polaritons in the vertical and in-plane directions. Moreover, the ultra-low-loss polaritons in α-MoO3 have been experimental measured [61, 190]. The polaritons are measured by scattering-type scanning near-field optical microscopy, which is shown in Fig. 9(d). The polariton amplitude lifetimes is larger than that of the plasmon polaritons in graphene and phonon polaritons in isotropically engineered boron nitride [61]. The hyperbolic in-plane dispersion is measured in Fig. 9(e), which is obtained by the Fourier transform of the probe measured near-field image [61].

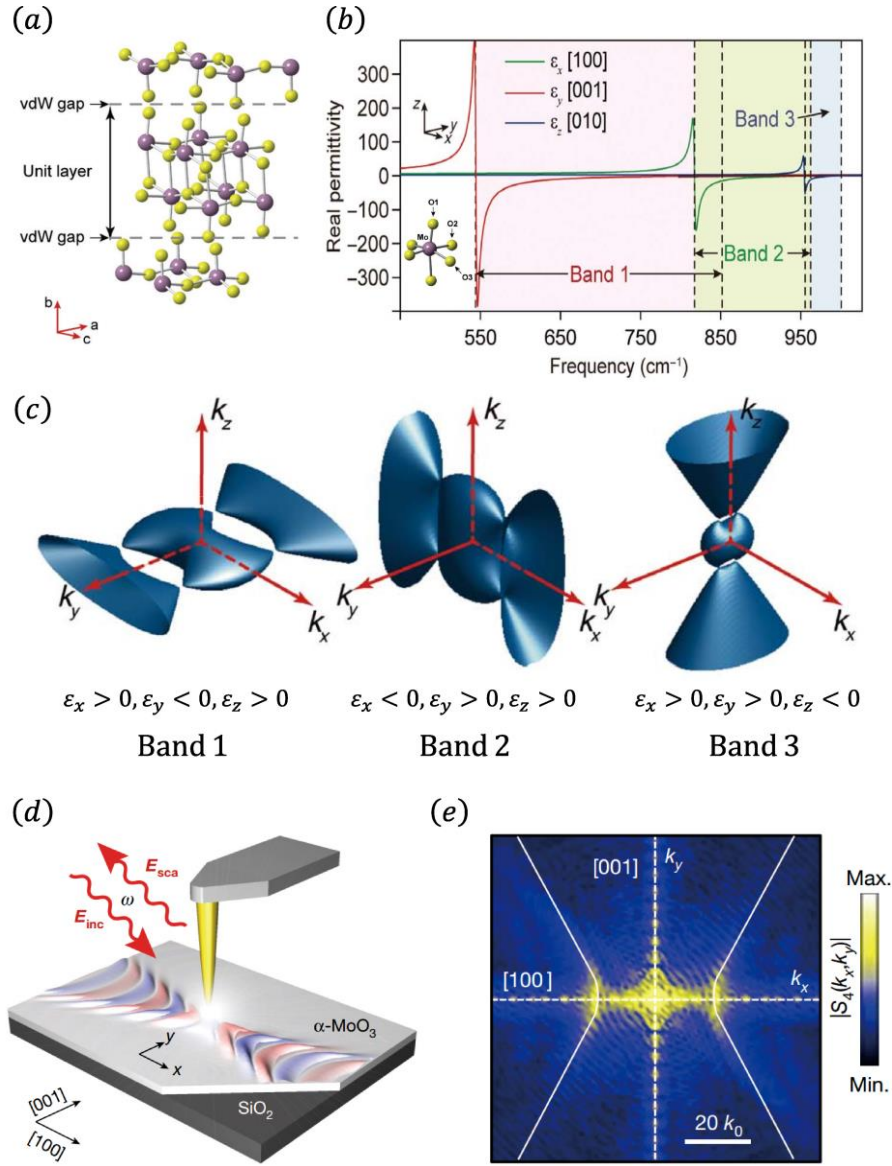

FIG. 9. The α-MoO$_3$ hyperbolic configuration. (a) Scheme of the α-MoO$_3$ crystalline. Reproduced with permission from Adv. Mat. 30, 1705318 (2018). Copyright 2018, Wiley-VCH. (b) The real part of the permittivity tensor of α-MoO$_3$. Three Reststrahlen bands are shaded in pink, yellow and blue, respectively. The inset graphic shows the unit cell of α-MoO$_3$, in which three different oxygen sites are indicated by O1–O3, respectively. (c) 3D IFCs have the biaxial hyperbolic dispersion in α-MoO$_3$. Reproduced under the terms of a Creative Commons AttributionNonCommercial-NoDerivs 4.0 International License. Sci. Adv. 5, eaav8690 (2019). Copyright 2019, The Authors. (d) Schematic of the scattering-type scanning near-field optical microscope in α-MoO$_3$ flake. (e) The hyperbolic in-plane dispersion is presented by Fourier transform of the near-field images at the frequency 893 cm$^{-1}$. The corresponding calculated IFC is shown by the white solid line. Reproduced with permission from Nature 562, 557 (2018). Copyright 2018, Springer Nature.

## 3. Topological transition of dispersion

### 3.1 Dispersion transition from closed ellipsoids to open hyperboloids

#### 3.1.1 Materials dispersion steered topological transition

Manipulating the topological property of IFC will provide a unique control for the interaction between light and matter, such as increased rates of spontaneous emission and negative refraction. A simplest topological transition of IFC from an elliptical dispersion to a hyperbolic dispersion is fulfilled by varying the frequency to make the real part of permittivity or permeability in dispersive metamaterials change its signs. So far, most of hyperbolic dispersions have been implemented in this way, including the metal/dielectric multilayer, metal nanowire array, circuit system and even the natural hyperbolic materials.

For the metal/dielectric multilayer and metal nanowire array, the permittivity of metal can be described by a Drude model:

$$\varepsilon(\omega) = \varepsilon_\infty - \omega_o^2 / (\omega^2 + i\omega\gamma) , \qquad (15)$$

where $\varepsilon_\infty$ is the high-frequency permittivity. $\omega_0$ and $\gamma$ denote the plasma frequency and damping frequency, respectively. At the low frequency, $\varepsilon(\omega)$ is negative with a large absolute value. As the frequency increases, $\varepsilon(\omega)$ gradually tends to zero and then becomes positive. As a results, based on Eqs. (9) and (10) the sign of the effective anisotropic permittivity can be tuned and the electric topological transition of IFC can be realized in this process. For the circuit system with lumped elements, both the permittivity and permeability are frequency dependent [163, 164]:

$$\varepsilon(\omega) = \varepsilon_a - \alpha/\omega^2, \; \mu_\perp(\omega) = \mu_b - \beta_x/\omega^2, \; \mu_{//}(\omega) = \mu_b - \beta_y/\omega^2 , \qquad (16)$$

where $\varepsilon_a$, $\mu_b$, $\alpha$, and $\beta_j$ ($j = x \text{ or } y$) are constants. We can clearly see that the sign of $\varepsilon(\omega)$, $\mu_\perp(\omega)$, and $\mu_{//}(\omega)$ are dependent on the frequency. So, the magnetic topological transition of IFC can be easily realized by tuning the frequency.

The topological transition also can be realized in the natural materials based on the materials dispersion. Take the vdW materials as an example, the principal components of the permittivity tensor of α-MoO$_3$ can be described by the following Lorentz equation [191, 192]:

$$\varepsilon_i(\omega) = \varepsilon_\infty^i (1 + \frac{\omega_{LO}^{i\,2} - \omega_{TO}^{i\,2}}{\omega_{TO}^{i\,2} - \omega^2 - i\omega\Gamma^i}), \tag{17}$$

where $\varepsilon_\infty^i$ is the high-frequency permittivity, and $\omega_{LO}^i$ and $\omega_{TO}^i$ denote the longitudinal and transversal optical phonon frequencies, respectively. $\Gamma^i$ means the damping factor. From Eq. (17), the signs of three principal components of the permittivity tensor can be tuned by changing the frequency. As a result, the topological transition of IFC can be realized. Therefore, changing the frequency to achieve the topological transition from ellipse to hyperbola is a usual way in many hyperbolic systems.

### 3.1.2 Loss induced topological transition

Intrinsic loss has long been thought to be a deteriorative factor in wave propagations. For example in the laser system, losses need to be overcome by a sufficient amount of gain to reach the threshold of lasing. Intuitively, the loss-induced suppression and revival of lasing have been proposed in the non-Hermitian system with an exceptional point [193]. The loss-induced transparent [194] indicates that the intrinsic loss could be an important and beneficial parameter and we need to re-examine the role of loss in electromagnetic wave manipulation. Usually, the topological transitions are realized by changing the frequency. Can the topological transition be realized at a fixed frequency by tuning another parameter? It is interesting that the parameter can be the intrinsic loss. For metamaterials in which the real part of the permittivity or permeability vanishes, the role of loss is highlighted and can even lead to a topological transition of dispersion from a closed elliptic curve to an open hyperbolic curve. The loss can enhance transmission and lead to collimation in the epsilon-near-zero (ENZ) metamaterials that have been theoretically proposed in metal-dielectric stacks [16, 17] and experimentally demonstrated in a circuit-based metamaterial [19]. Based on this method, the wave-front manipulation and energy collimation also have been systematically studied in the metal-line system [195]. Specially, beyond the ENZ metamaterials with a very flat elliptical dispersion, loss-induced topological transition of dispersion can occur in metamaterials with an arbitrary positive real part of permittivity and permeability [143]. We suppose that $\mu_{//}$ is a complex number while $\mu_{\perp}$ is a real number. Considering the boundary condition, the real and imaginary parts of $k_z$ along optical axis direction can be obtained by:

$$k_z^2 = \mu_{\perp}[\varepsilon k_0^2 - (k_x^2 + k_y^2)/\mu_{//}]. \tag{18}$$

Supposing $\varepsilon = \mu_{\perp} = 1$, and $\mu_{//} = 0.3 + \text{Im}(\mu_{//})i$, Fig. 10(a) shows how the closed ellipsoid evolves into the open hyperboloid by changing the loss at a fixed frequency. The cyan solid and red mesh surfaces respectively give the relationship of real part and imaginary part with $k_x$ and $k_y$. For the case without loss ($\text{Im}(\mu_{//}) = 0$), the IFC corresponds to a closed ellipsoid, as is shown in Fig. 2(a). However, as the loss increases to $\text{Im}(\mu_{//}) = 0.05$, the closed ellipsoid will gradually break at the center part (see the left graph of Fig. 10(a)). When the loss further increases to $\text{Im}(\mu_{//}) = 0.2$, the ellipsoid is totally broken and evolves into an open hyperbolic-like curve (see middle graph of Fig. 10(a)). If we continue adding the loss, the hyperbolic curve becomes more and more flat and the propagation loss becomes smaller and smaller (see right graph of Fig. 10(a)). The change of the IFC will strongly modify the propagation properties of electromagnetic wave. When the IFC becomes flat, the electric fields are collimated. The corresponding measured emission patterns of a source in the circuit metamaterials are shown in Fig. 10(b). As loss increases, the measured emission patterns change from the two connected sectors to a column [143]. Figure 10(c) presents one sample of the circuit metamaterial, in which the loss can be flexibly controlled by the resistors. By measuring the voltage pattern, loss-induced field enhancement and collimation have been observed in Fig. 10(d). The topological transition does not depend on the real part of $\mu_{//}$, as is shown in Fig. 10(e). It is seen that for an arbitrary real part of $\mu_{//}$, the topological transition will occur so long as the imaginary part of $\mu_{//}$ is large enough. Therefore, without using gain to compensate the loss [196, 197], tuning the anisotropic loss is also an effective way to control the dispersion.

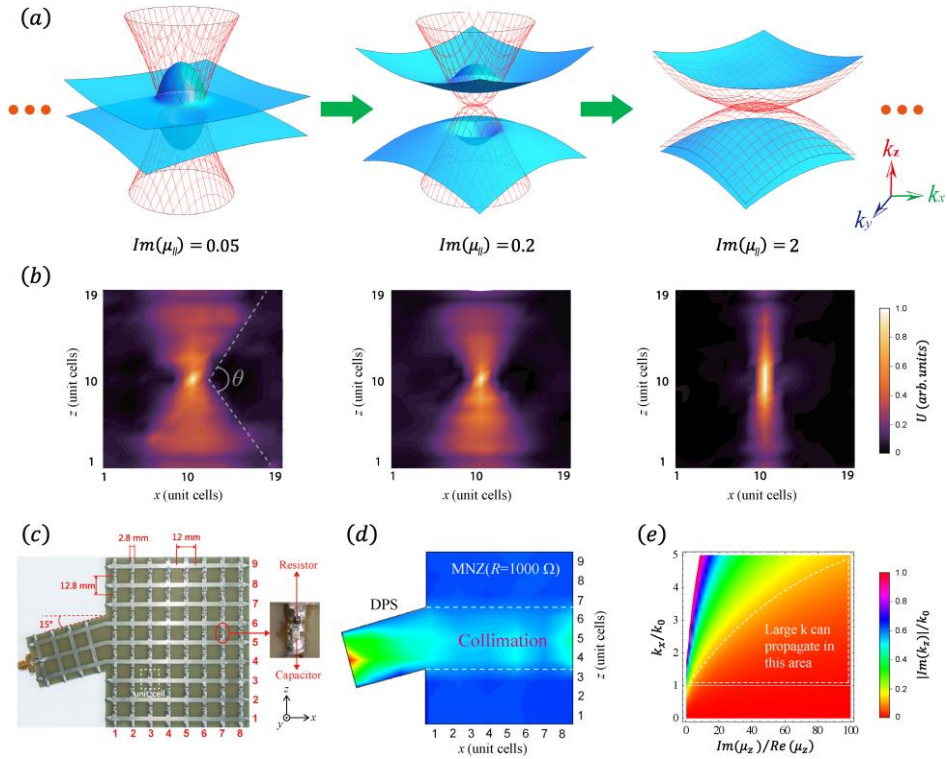

FIG. 10. (a) Topological transition of IFC controlled by the imaginary part of $\mu_{//}$. (b) The measured emission patterns for different values of resistors at a fixed frequency. Reproduced with permission from J. Appl. Phys. 119, 203102 (2016). Copyright 2016, AIP Publishing. (c) The circuit based metamaterials with a wedge to demonstrate the collimation. (d) The measured voltage distributions of (c) at 1.56 GHz. (e) The normalized imaginary part of $k_z$ as a function of the $Im(\mu_z)/Re(\mu_z)$ and $k_x/k_0$ when $\varepsilon = \mu_x = 1$. Reproduced with permission from Phys. Rev. B 91, 045302 (2015). Copyright 2015, American Physical Society.

### 3.1.3 Actively controlled topological transition

The topological transition of dispersion in the passive systems can be realized by changing the frequency or adding anisotropic loss. However, the ability to actively tune these effects remains elusive, and the related experimental observations are highly desirable. A very interesting topic is how to realize a topological tuning of IFC in an active manner because the actively tunable IFC is very useful in the design of new active devices. Recently, a novel implementation of topological transition is proposed based on the 2D semiconductors such as graphene. Varying the graphene chemical potential by using the gate voltage yields an effective manner to control the graphene conductivity in the terahertz and infrared frequencies [198-200]. The combination of metamaterials and graphene provides an important way for people to actively manipulate electromagnetic waves, such as graphene-based metasurface [201, 202], subwavelength focusing [203], negative refraction [204], and slow light [205]. Schematic of the graphene/dielectric multilayer structure is shown in Fig. 11(a). The dielectric layers are marked in cyan and the graphene layers are shown by the single layer of carbon atoms. Effective permittivity of graphene sheet is characterized by its surface conductivity $\sigma(\omega, \mu_c)$ as [198]:

$$\varepsilon_g = 1 + i\sigma(\omega, \mu_c)/\varepsilon_0 \omega t_g, \tag{19}$$

where $\mu_c$ and $t_g$ denote the chemical potential and thickness of graphene sheet. Specially, $\mu_c$ can be tuned by the gate voltage. Therefore the permittivity of graphene sheet can be tuned by changing the value of $\mu_c$. At the certain frequency region, $\varepsilon_g$ is a large negative value. This means the graphene has metallic properties. The permittivity and thickness of dielectric layer are 2.25 and 50 nm, respectively [204]. According to Eq. (9), the anisotropic permittivity tensor of the graphene/dielectric multilayer structure can be obtained. The value and the sign of $\varepsilon_{//}$ can be flexibly adjusted by changing the value of $\mu_c$. As the value of $\mu_c$ increases from 0.1 to 1 ev, the sign of $\varepsilon_{//}$ will reverse at 0.54 ev while the sign of $\varepsilon_{\perp}$ remains positive in Fig. 11(b). The hyperbolic regime and the elliptical regime are marked in Fig. 11(b). In addition, based on the array of graphene strips, researchers actively tune the dispersion from an elliptic curve to a hyperbolic-like curve by changing the external gate voltage [84, 85]. The corresponding dispersion relation equation is written as [84]:

$$\eta_0^2(k_x^2\sigma_{xx}+k_y^2\sigma_{yy})^2(k_x^2+k_y^2-k_o^2)-4k_o^2(k_x^2+k_y^2)^2=0, \qquad (20)$$

where $\eta_0$ denotes the impedance of the free space. Effective conductivity tensor in the HMS consisting of a graphene strip array is presented in Fig. 11(c). When the condition $Im(\sigma_{xx})\,Im(\sigma_{yy})<0$ is satisfied, the hyperbolic dispersion can be realized. The topological transition from a closed ellipse to an open hyperbolic dispersion is associated with the change of condition from $\text{Im}(\sigma_{xx})<0, \text{Im}(\sigma_{yy})<0$ to $\text{Im}(\sigma_{xx})\,\text{Im}(\sigma_{yy})<0$. The propagating direction of SPP in the HMS with graphene is controlled by the applied $\mu_c$. Figures 11(d) and 11(e) are the distributions of electric field magnitude when the surface plasmons are excited for $\mu_c=0.1$ and 0.3 ev, respectively.

In high frequency region, besides the difficulty in fabricating the multilayers, it is not easy to precisely add the same external voltage to each graphene layer [206-208]. In microwave regime, circuit loaded with variable capacitance diode provide a flexible platform to experimentally demonstrate the actively controlled topological transition. In these circuits, the value of capacitance can be actively tuned by an external voltage. In 1D circuit systems, the inverse, zero Doppler effect and the frequency mixing have been observed[209-211]. The 2D circuitsample used to achieve the topological phase transition is shown in Fig. 11(f). A direct voltage source is connected to the sample from the top, which is marked by the red arrow. The signal is input at the center of the sample [144].

Based on Eq. (14), the anisotropic permeability of the circuit based metamaterial can be tuned by changing the capacitance. As the voltage increases, the capacitance of the varactor decreases accordingly. Therefore, the topological transition from elliptic dispersion to hyperbolic dispersion occurs. For example, when the U=1 V, the IFC is a closed elliptic curve and the source can propagate in all in-plane directions, as is shown in Fig. 11(g). However, when U=20 V, the IFC becomes an open hyperbolic curve and the source can only propagate in some directions, as is shown in Fig. 11(h). So the actively controlled topological transition of dispersion by changing the external voltages can be easily observed in the circuit based metamaterials.

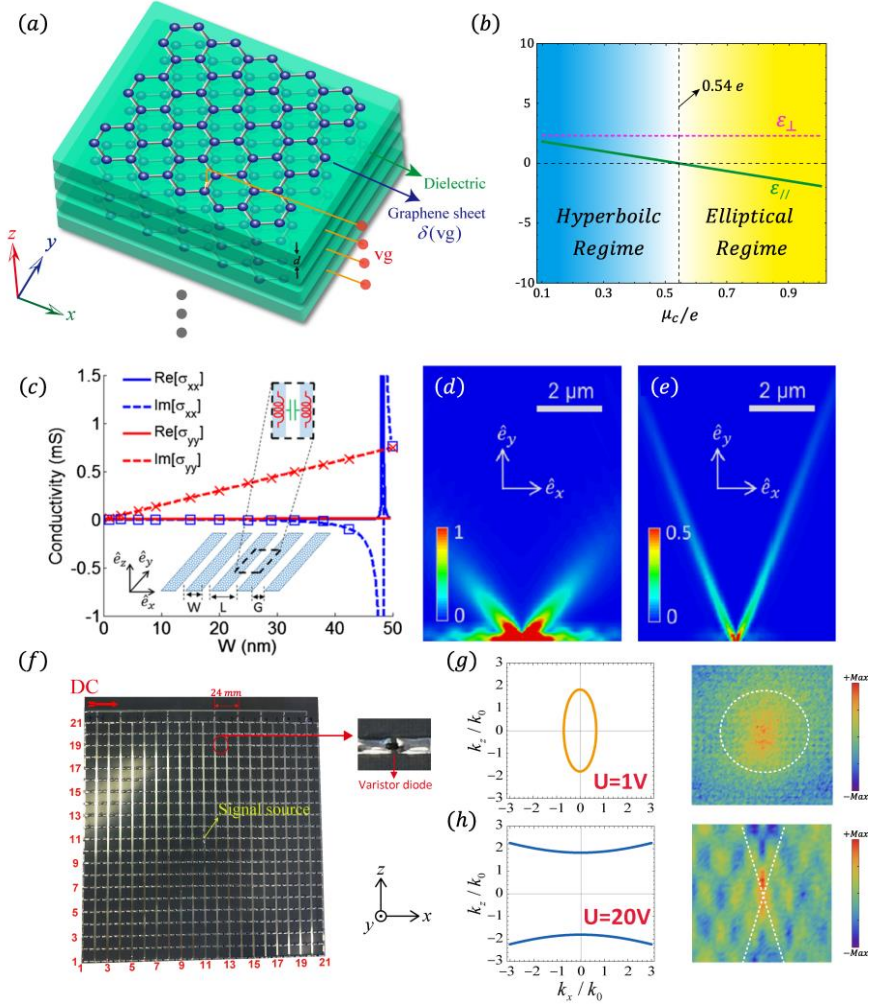

FIG. 11. (a) Schematic of the periodic structure composed of graphene/dielectric multilayers. The chemical potential of the graphene layer can be tuned by the external voltage, and under certain external voltage this multilayer structure has a hyperbolic dispersion. (b) Real parts of $\varepsilon_\perp$ and $\varepsilon_{//}$ based on the multilayer structure at a fixed frequency of 40 THz. (c) Effective conductivity tensor of the uniaxial HMS consisting of graphene strips. (d) When $\mu_c = 0.1$ ev, the distribution of the electric field for SPPs excited on the graphene HMS. (e) Similar to (d), but the graphene chemical potential is 0.3 ev. Reproduced with permission from Phys. Rev. Lett. 114, 233901 (2015). Copyright 2015, American Physical Society. (f) Circuit-based metamaterials in circuit system that can be actively tuned by an externally applied voltage. The insets show the enlarged lumped variable capacitance diode. (g) The IFC and the measured electric field $E_y$ when the U=1 V. (h) Similar to (g), but the voltage is U=20 V. Reproduced under the terms of a Creative Commons AttributionNonCommercial-NoDerivs 4.0 International License. Appl. Sci. 8, 596 (2018). Copyright 2018, The Authors.

### 3.2 Transition points at two kinds of topological transition

#### 3.2.1. Anisotropic zero-index metamaterials

By changing the sign of $\mu_{//}$ from a positive value to a negative value while the sign of $\varepsilon$ and $\mu_\perp$ remain positive, the topological transition will occur, as illustrated in Fig. 12(a). In this process, when $\mu_{//}$ changes from a positive value to $\mu_{//} \to 0$, the IFC changes from an ellipsoid to a very flat ellipsoid. Then, after the sign of $\mu_{//}$ changes to a negative value, the IFC changes to an open hyperboloid. At the transition point ($\mu_{//} \to 0$), the material is the anisotropic mu-near-zero (AMNZ) material whose IFC is a very flat ellipsoid. Although the findings are based on the magnetic-type hyperbolic topological transition, similar results can be extended to the electric-type hyperbolic topological transition when the permittivity and permeability are exchanged. Atthe transition point of the electric-type topological transition, the material is the anisotropic epsilon-near-zero (AENZ) material. At the transition point, many unusual transportation properties of light in AENZ and AMNZ metamaterials have been demonstrated, such as spatial power combination [212], bending waveguide [213-215], field shielding [216], total reflection [217] and perfect absorption [218]. Interestingly, it has been found that flux manipulation can be achieved in the anisotropic zero index metamaterials [219-222]. When the scatters are placed in the background material, the inhomogeneity profile $f(x, z)$ is a spatially varying function. For the ordinary background material (such as air), the incident light will be scattered by different scattering objects. Since the scattered waves distribute energy flux in all directions, transmission in the forward direction is reduced. However, the scattering phenomenon will be strongly modified by the AENZ metamaterial ($\varepsilon_x \to 0^+$) in Fig. 12(b). In this case, the scattered waves are the evanescent waves in the horizontal direction instead of propagating waves in all directions. The physical mechanism can be understood from the IFC in Fig. 12(c). The cyan circle and orange ellipse correspond to the IFC of air and AENZ media, respectively. $K_\perp$ denotes the wavevector of scattered wave in the $z$ direction. When the scattering occurs in the air background material, $K_\perp$ still fall within the maximum value of the IFC of air in Fig. 12(c), the scattered wave thus still be the propagating wave along various directions. However, when the scattering occurs in the AENZ background metamaterial, $K_\perp$ will exceed the IFC of the AENZ in Fig. 12(c) and the evanescent waves will be excited. This scattering behavior directly results in completely different scattering propagation phenomena. It is this unusual scattering phenomenon that enables a new way to regulate the electromagnetic energy flow at the sub-wavelength scale. Owing to the excitation of the evanescent scattered waves in this inhomogeneous and anisotropic ZIMs, the electromagnetic energy flux can be controlled according to the spatial profile of the nonzero $\varepsilon$ or $\mu$ component, as are shown in Figs. 12(d) and 12(e). This energy flow manipulation in AMNZ metamaterial have been experimentally demonstrated in Fig. 12(f). The sample is constructed based on the circuit system with lumped elements. The measured field distributions are shown in Fig. 12(g), where the $E_y$ is uniform in the $z$ direction. However, compared with the background medium, $H_z$ in the region of scatter is enhanced and $H_z$ is excited with the evanescent waves. Figure 12(g) shows the magnetic components can be controlled in the subwavelength scale in the AMNZ metamaterial.

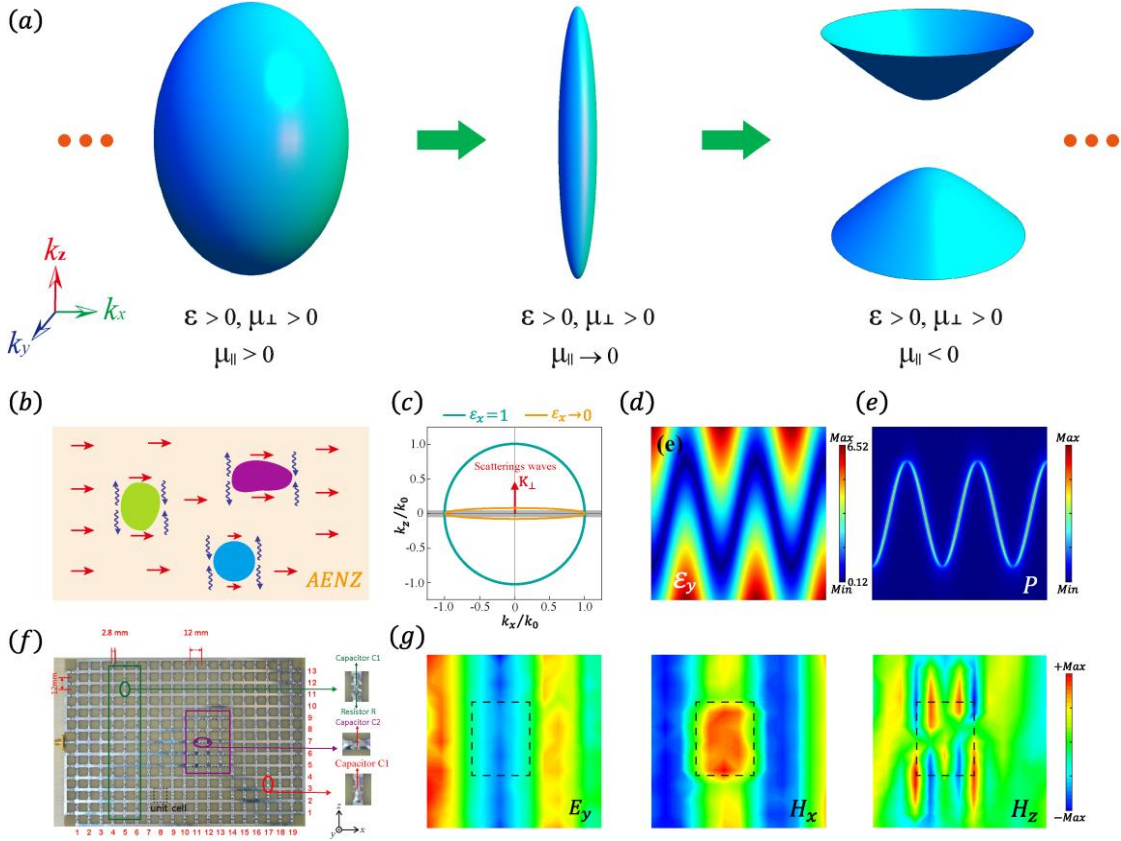

FIG. 12. (a) Magnetic topological transition of IFC from the closed ellipsoid to an open hyperboloid when $\mu_{//}$ changes from a positive value to a negative value while $\varepsilon$ and $\mu_\perp$ remain unchanged. (b) Schematic of electromagnetic wave propagation and scattering in an AENZ system with defects. (c) IFCs of the air and the AENZ background media. $K_\perp$ denotes the perpendicular component of the wavevectors of the scattered waves. (d), (e) Free routing of energy flux in a designed sinusoidal path when electromagnetic wave propagates in the direction of $\varepsilon_x \to 0$. Reproduced with permission from Phys. Rev. Lett. 112, 073903 (2014). Copyright 2014, American Physical Society. (f) Circuit-based AMNZ metamaterials in circuit system where the defect region is marked by the purple rectangle. (g) The measured field distributions of $E_y$, $H_x$, and $H_z$ from column 7 to 19 in the sample (f). Reproduced with permission from Europhys. Lett. 113, 57006 (2016). Copyright 2016, Institute of Physics.

### 3.2.2 Linear crossing metamaterials

From Eq. (3), we can clearly see that the IFC of magnetic HMMs is not only related to the anisotropic permeability, but also to the permittivity. Normally, the topological transition of two kinds of electric hyperbolic metamaterials will take place when the sign of anisotropic permeability are exchanged and the permittivity is maintained at 1. However, a new kind of linear crossing dispersion will exist when the sign of permittivity changed while the sign of anisotropic permeability remains unchanged, as is shown in Fig. 13(a). Specially, in a 2D case, when $\mu_\perp \mu_{//} < 0$ and $\varepsilon \to 0$ the IFC at the transition point from metal-type HMM to dielectric-type HMM (or vice versa) corresponds to two intersecting lines. This material with two intersecting linear dispersions can be called linear-crossing metamaterial (LCMM). Because the signs of in-plane permeability are opposite and the permittivity tends to zero, LCMM may simultaneously possess the characteristics of HMM and zero-index media. On the one hand, similar to HMM, high-$k$ waves can be supported by the LCMM because of the open IFC. In addition, the positive and negative refractions can be controlled by the sign of in-plane permeability just like in the HMM. The 2D IFC of the LCMM with negative refraction is shown in Fig. 13(b). On the other hand, the group velocity and phase velocity in LCMM are perpendicular to each other, which leads to the zero phase accumulation along the propagation path just as in a zero-index medium. Interestingly, LCMM can possess many unique properties [22]. For example, the directional propagation can be observed when the light is obliquely incident to the LCMM, as is shown in Fig. 13(c). Different from the negative refraction in double negative metamaterial, the refracted angle in LCMM is independent of the incident angle. On the other hand, the novel beam splitting phenomenon [223, 224] will occur in LCMM when the light is normally incident on the structure, as is shown in Fig. 13(d) [22]. Owing to the fact that the refracted light is locked in the two fixed directions in LCMM, the propagation of light will not be affected by the inner defect so long as it is not placed in the path of refraction. Based on circuit system, two kinds of LCMMs respectively with positive and negative refraction are constructed. And the super-resolution imaging with partial cloaking have been experimentally demonstrated in Fig. 13(e). The measured distribution of normalized electric field is shown in Fig. 13(f). The region of defect is marked by the yellow dashed rectangle and the positions of images are marked by the two red dashed circles. Moreover, to quantitatively demonstrate the subwavelength imaging with partial cloaking, the normalized intensity of electric fields at the exit plane in two cases with and without defect are shown in Fig. 13(g). From Figs. 13(f) and 13(g), the super-resolution imaging with partial cloaking based on the effective LCMMs have been experimentally demonstrated. Recently, the omnidirectional invisibility cloak from a wide range of incident angles also has been discovered in the anisotropically doped ENZ metamaterial, which is transparent to any incident illumination [225].

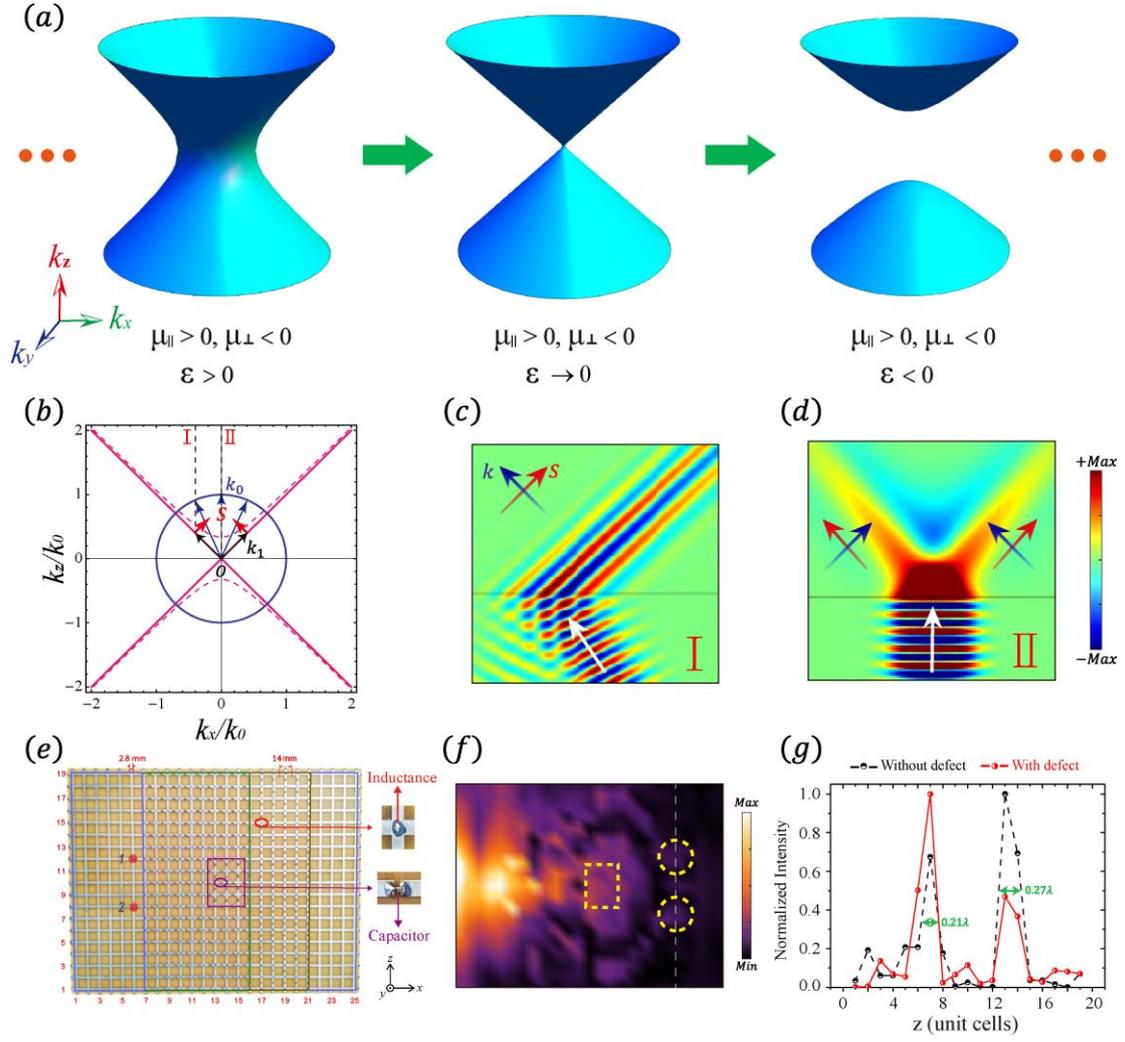

FIG. 13. (a) Magnetic hyperbolic topological transition of IFC from the metal-type hyperbolic dispersion to dielectric-type hyperbolic dispersion when $\varepsilon$ changes from a positive value to a positive value while $\mu_{//}$ and $\mu_{\perp}$ remain unchanged. (b) The IFC of the LCMM with $\mu_z=1$, $\mu_x=-1$ and $\varepsilon \to 0$. In this LCMM, a negative refraction independent of the incident angles will take place at oblique incidence (c). However, at normal incidence the incident light will split along two directions (d). (e) Circuit-based LCMMs. Two sources are marked by red dots. (f) Measured distribution of normalized electric field. Two images are marked by the yellow circles. (g) Normalized field intensity distributions along the $z$ direction at the exit plane. Reproduced with permission from Phys. Rev. Appl. 10, 064048 (2018). Copyright 2018, American Physical Society.

## 4. Dispersion control in Hypercrystals

### 4.1 Controlling the dispersion of band structure

Because of the unusual properties of HMM, it is expected that a composite structure with HMMs could also has an unprecedented ability to manipulate the electromagnetic waves. Recently

E. E. Narimanov has studied a structure in which one dielectric component of a one-dimensional (1D) photonic crystal (PC) is replaced by a HMM component and he calls such structure a photonic hypercrystals (PHC) [95]. In a PHC, the special dispersion of HMM would strongly modify the Bragg scattering and thereby manipulate the bandgaps of the structure, as we will see later.

The most remarkable feature of PCs is that they have photonic band gaps, which can be used to suppress spontaneous emission [226] and localize photons [227]. In particular, an omnidirectional gap based on 1D PCs is very useful in many application such as optical reflectors at all angles [228]. For the traditional dielectric PCs $(AB)_n$, the band gaps come from the Bragg scattering mechanism. At the Bragg frequency of the first band gap, the Bragg condition is written as:

$$\phi = (k_{Az}d_{Az} + k_{Bz}d_{Bz})|_{\omega Brg} = m\pi, \qquad (21)$$

where $k_{Az}$ and $k_{Bz}$ are the wave wavevectors in the propagating direction in medium $A$ and $B$, respectively. $\phi$ denotes the propagating phase in a unit cell. The dispersion relation of a dielectric is $k_x + k_z = \varepsilon\mu(\omega/c)^2$, where $k_x = k_0 \sin\theta$ is parallel wavevector. When the incident angle $\theta$ increases, $k_x$ will increase while $k_{Az}$ and $k_{Bz}$ decrease. If the frequency is still $\omega_{Brg}$, the Bragg condition in Eq. (21) will no longer hold. Therefore, to maintain the Bragg condition, the frequency should increase. Therefore, as the incident angle increases, the gap will shift toward high frequency. In other words, in the traditional 1D PC, the gap edges change with the incident angles and the gap would even close at Brewster angle. To overcome this problem, the omnidirectional gaps independent of the incident angles are proposed in special 1D PCs, including the average index (zero-$\bar{n}$) gap under the mechanism of the phase cancellation of propagating waves between positive- and negative-index materials [229-233] and zero effective phase (zero-$\phi_{eff}$) gap under the mechanism of the compensation of the exponentially increasing and decreasing waves in two kinds of single-negative materials [234-237].

The PHC provides a new physical mechanism for designing the angle-independent (dispersionless) gap. The schematic of a 1D PHC is shown in Fig. 14(a). Suppose that $A$ layer is hyperbolic medium while $B$ layer is dielectric. In Fig. 14(b), it is seen that the slopes of isofrequency curves of dielectric and hyperbolic medium have opposite signs. When the incident angle increases, $k_x$ will increase. From the isofrequency curve, when $k_x$ increases, $k_{Bz}$ or the phase will decrease, but $k_{Az}$ or the phase will increase. Therefore, as the incident angle increases, the phase variations in two media have opposite signs. It is this phase variation compensation (PVC) effect that leads to the dispersionless gap. Please note that it is PVC effect rather than phase compensation effect because the total phase is still $\pi$. From Eq. (21), the condition of the dispersionless gap is [238]:

$$\frac{\partial \Phi}{\partial k_x} = (d_A \frac{\partial k_{Az}}{\partial k_x} + d_B \frac{\partial k_{Bz}}{\partial k_x})\Big|_{\omega_{Brg}} = 0 . \tag{22}$$

The PVC condition of the dispersionless gap can be analytically approximated in the $\varepsilon_B \gg 1$ and $|\varepsilon_{Az}| \gg 1$ limit as [238]:

$$d_A = \frac{\pi c}{\omega_{Brg}} \frac{1}{\sqrt{\varepsilon_{Ax}}(1-\frac{\varepsilon_B}{\varepsilon_{Az}})}, d_B = \frac{\pi c}{\omega_{Brg}} \frac{1}{\sqrt{\varepsilon_B}(1-\frac{\varepsilon_{Az}}{\varepsilon_B})} . \tag{23}$$

The phase compensation effect also can be clearly explained in Fig. 14(c). Based on the PVC condition, the reflection spectra of 1D PHC ([TiO$_2$/HMM]$_3$) for TM waves is calculated in Fig. 14(d). The HMM is realized by two subwavelength unit cells ([TiO$_2$/Ag]$_2$). The corresponding measured reflectance spectra at the incident angles are shown in Fig. 14(e). It is seen that two band edges (reflective dips near the band gap) vary very little as the incident angle changes from 10º to 40º. The measured gap edges are marked in Fig. 14(d) and we see that the experimental results coincide well with the simulated results [239]. In addition to the dispersionless band gap, the red shift of bandgap can also be realized by $\partial \Phi / \partial k_x > 0$, which is distinct from the blue shift of bandgap for conventional dielectric PCs. The simulated and measured red shift bandgaps for TM waves are shown in Figs. 14(f) and 14(g), respectively [240]. It should be pointed out that, for TE ordinary waves, the subwavelength metal/dielectric multilayer structure is just like a dielectric. As a result, the bandgap is blue shifted, as is shown in Fig. 14(f).

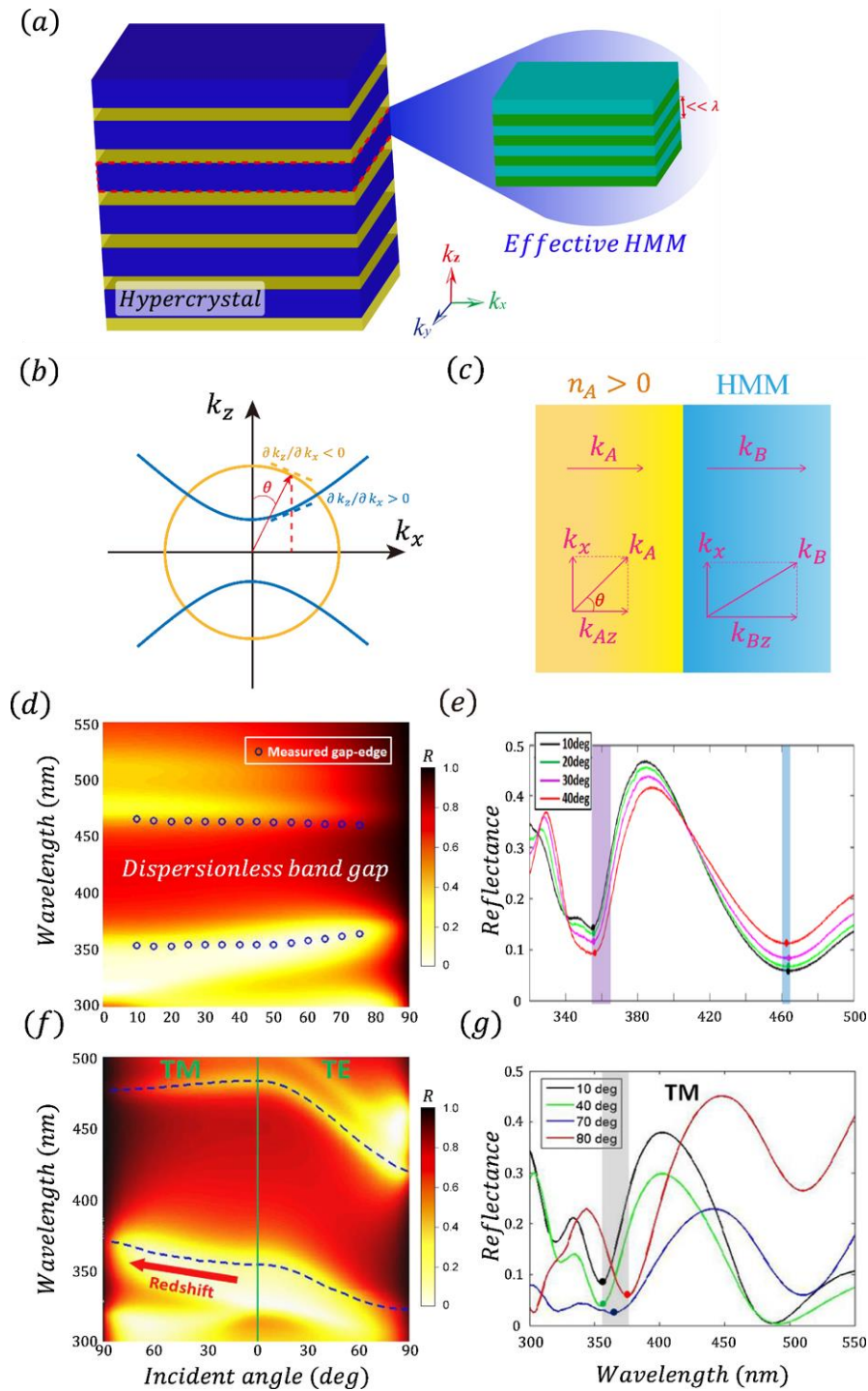

FIG. 14. (a) Schematic of the 1D PHC composed of a HMM layer and a dielectric layer. The enlarge graph denote that the HMM layer is mimicked by a metal/dielectric layered structure. (b) IFCs of the HMM and dielectric media in (a) are shown by the blue and orange lines, respectively. (c) Schematic of a plane wave propagating in one unit cell of the 1D PHC. (d) Reflection spectrum of 1D PHC ([TiO$_2$/HMM]$_3$) for TM waves. The HMM is realized by two subwavelength unit cells ([TiO$_2$/Ag]$_2$). (e) Measured reflectance at the incident angles of 10º, 20º, 30º, and 40º, respectively. Reproduced with permission from Appl. Phys. Lett. 112, 041902 (2018). Copyright 2018, AIP Publishing. (f) Similar to (d), but for two kinds of polarizations and the thickness of metal and dielectric layers is adjusted slightly. (g) Measured reflectance at the incident angles of 10º, 40º, 70º, and 80º, respectively. Reproduced with permission from Phys. Rev. Appl. 10, 064022 (2018). Copyright 2018, American Physical Society.

## 4.2 Cavity modes and edge modes with special dispersion

The dispersionless gaps can be used to realize wide-angle cavity mode. The quality factor for the dispersionless cavity mode hardly changes when the plane waves launch at different angles, which is very useful when the cavity mode is excited with finite-sized sources. Schematic of the 1D PHC with dielectric defect marked in pink is shown in Fig. 15(a). It is seen from Fig. 15(b) that the defect or cavity mode inside the dispersionless gap remains nearly invariant with incident angles, owing to the PVC effect [238].

Moreover, in a heterostructure composed of a 1DPC and metal (Fig. 15(c)), an edge mode would occur at the interface of the 1DPC and the metal [241]. The phase matching condition of edge states can be expressed by [242, 243]:

$$\phi_M + \phi_{pc} = 0, \quad \phi_M, \phi_{PC} \in (-\pi, \pi), \tag{24}$$

where $\phi_M$ and $\phi_{pc}$ denote the reflection phases of metal layer and 1DPC, respectively. From this condition for the edge mode, the reflection phase of the metal changes little with the incident angle. This means that the dispersion properties of edge mode are essentially determined by that of the 1DPC. So, a nearly dispersionless edge mode can be obtained in a heterostructure composed of a metal layer and 1D PHC because of the dispersionless gap of the 1D PHC. At this nearly dispersionless edge mode, the light can enter the heterostructure and is fully absorbed by the lossy metal. The measured absorption peaks marked by white circles and the simulated absorption spectrum are shown in Fig. 15(d) [244]. T The experimental results agree well with the simulated results. The absorption peak is always near one in a wide range of incident angles. Therefore, based on the dispersionless gaps, the wide-angle perfect absorbers can be designed [241].

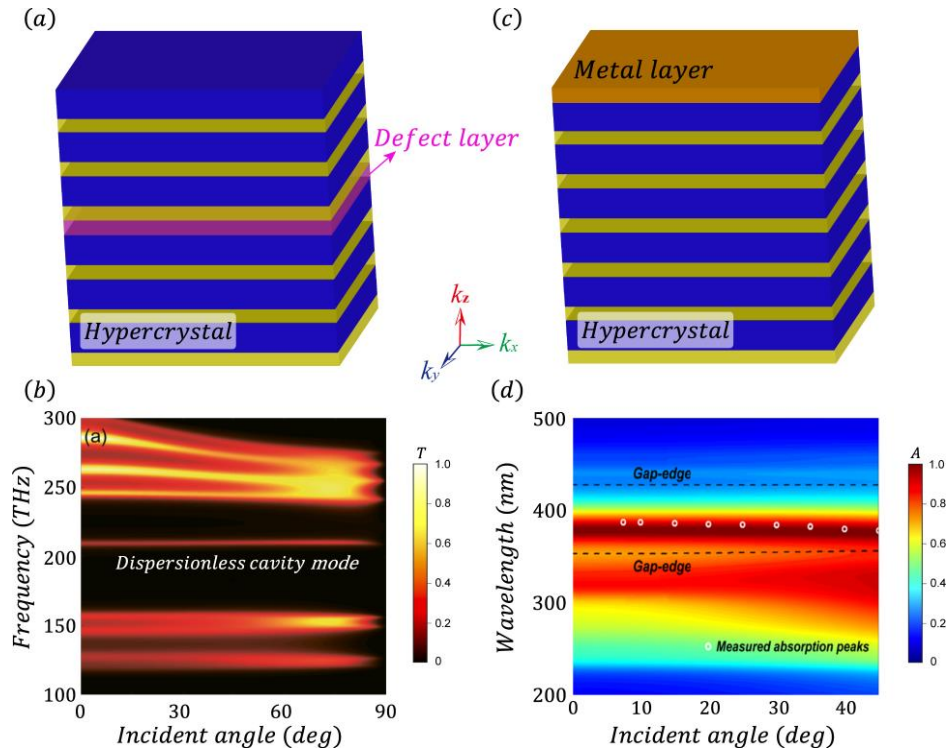

FIG. 15. (a) Schematic of the 1D PHC with a defect layer at the center of the structure. (b) Transmittance spectrum of (a) for all angles of incidence. Reproduced with permission from Phys. Rev. B 93, 125310 (2016). Copyright 2016, American Physical Society. (c) Schematic of the heterostructrue composed of a metal layer and a 1D PHC.(d) Absorption spectrum of (c) from 0º to 45º. Some measured absorption peaks are marked by the white hollow circles. Reproduced with permission from Opt. Express 27, 5326 (2019). Copyright 2019, OSA Publishing.

Recently, based on the dispersion control, PHC also have been proposed to realize the Dirac dispersion [245], to design wide-angle biosensors [246], and to construct the Veselago lens [247]. In addition to the 1D PHC, 2D PHC also has been proposed. And it can greatly enhance the light-matter interaction, such as increasing the spontaneous emission [248-250], and design accidental degenerate double Dirac cones [251, 252].

# 5. Applications

## 5.1 Hyperlens

In imaging optics, the information of the objects is carried by the scattering field with various wavevectors. Specially, the large-feature (fine details) is carried by the propagating waves (evanescent waves) with small (large) wavevectors. For the conventional lens, the evanescent waves will confined at the near-field region and decay exponentially with transmission distance, as is shown in the upper row of the Fig. 16(a). The missing sub-wavelength information results in the diffraction limit of imaging. By using the evanescent field amplification effect, the diffraction limit can be overcome and the imaging quality can be greatly improved [125, 253, 254]. In this case, owing to the coupling of the interface modes, the near-field components can be recovered at image plane. According to this physical principle, a silver film with $\varepsilon = -1$ is used to achieve sub-wavelength imaging, which verifies the potential application of superlens [253].

In addition to the superlens based on the near-field amplification, hyperlens based on the conversion of near fields into far fields can also realize the sub-diffraction-limited resolution [45, 46]. One of the most fascinating features of HMM is that it supports large-$k$ modes because of the open hyperboloid dispersion. HMMs can convert the near fields into high-$k$ propagating waves and thus overcome the diffraction limit, as is shown in the lower row of the Fig. 16(a). The cylindrical prototype of hyperlens is obtained by bending the flat layers (alternating metal/dielectric multilayers) into curved layers, as is shown in Fig. 16(b). The resolution and magnification in the hollow cylinder hyperlens can be estimated by the ratio between the inner and outer radii of the structure. The hyperlens beyond the diffraction limit also have been proposed for photolithography. It can be used to generate deep subwavelength patterns from diffraction-limited masks, as is illustrated in Fig. 16(c) [255, 256]. Besides, the hyperlens are realized by nano-metal wire array in the near-infrared and mid-infrared wavelengths [257, 258], and the quarter-wave image can be well observed in the image plane. Although hyperlens with the open hyperbolic dispersion can realize super-resolution imaging very well, the elliptical dispersion (such as the anisotropic zero index metamaterials) can also realize the superlens as long as the coverage for lateral wavevectors is large enough [259].

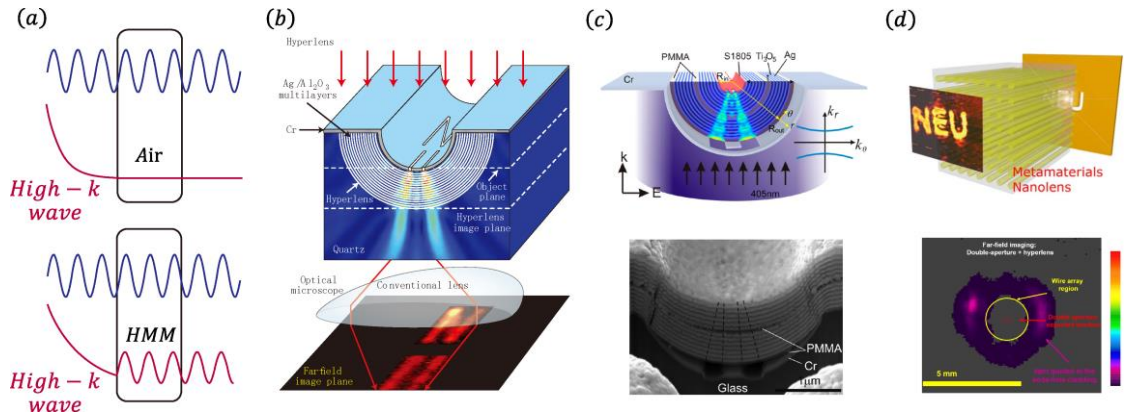

FIG. 16. (a) Schematic of the conventional lens and hyperlens. Reproduced with permission from Nat. Mater. 7, 435 (2008). Copyright 2008, Springer Nature. (b) Hemi-cylindrical hyperlens composed of metal-dielectric multilayers. Reproduced with permission from Science 315, 1686 (2007). Copyright 2007, American Association for the Advancement of Science. (c) Light photolithography with hyperlens. Reproduced with permission from ACS Nano 12, 542 (2018). Copyright 2018, American Chemical Society. (d) Hyperlens realized by nano-metal wire array and the far-field images. Reproduced with permission from Appl. Phys. Lett. 96, 023114 (2010). Copyright 2010, AIP Publishing. and Reproduced with permission from Opt. Express 27, 21420 (2019). Copyright 2019, OSA Publishing.

### 5.2 Long-range energy transfer

The interaction between two dipoles is a basic physical effect, which plays an important role in many phenomena such as Casimir force, van der Waals force and vacuum friction [260, 261]. The interaction between two dipoles is based on the near-field coupling mechanism. For a near field with a large wavevector component, it decays rapidly as the spatial distance increases. Therefore, in a vacuum or ordinary medium environment, the interaction distance between two dipoles is very short and it is generally much smaller than one wavelength. However, the emergence of hyperbolic materials as a new environment offers the possibility to overcome the limitation of short-range interaction in traditional environments. HMMs can support the large wave-vector modes. This means in HMMs the near fields can be converted into the high-$k$ propagating waves, as is shown in Fig. 17(a). Recent theoretical studies have shown that the interaction distance between two dipoles can be greatly increased to the order of one wavelength by using HMMs [50-53]. Figure 17(b) shows that two dipoles placed on HMS can realize long-range interaction [51]. $\theta_{xy}$ denotes the angle with respect to the optical axis of HMs. Along the resonance cone angle $\theta_{xy} = \tan^{-1}\sqrt{-\varepsilon_\perp/\varepsilon_{//}}$, the cooperative Lamb shift as a function of the separation distance is shown in Fig. 17(c). Inset in Fig. 17(c) shows giant enhancement of Förster resonance energy transfer [51].

HMM can also be utilized to greatly enlarge the coupling distance between a dipole-like bright atom and a quadrupole-like dark atom [54]. The electromagnetically induced transparency (EIT) in the meta systems like three-level system requires significant coupling between the bright atom and dark atom modes [262, 263]. In a conventional medium, the distance of near-field coupling between the bright and dark atoms has to be λ/20~λ/40, where λ is the operation wavelength. However, in a HMM, a long-range EIT in which the bright and dark atoms are separated far away is observed in a microwave experiment. The microwave HMM is constructed based on the circuit system with lumped elements, as is shown in Fig. 17(d). Compared to the ordinary materials, the coupling length can be enhanced by nearly two orders of magnitude in the HMM. Accordingly, the energy transfer from the bright atom to the dark atom will be enhanced greatly. The long-range energy transfer is shown in Fig. 17(e). The HMM-mediated EIT survives even when the coupling distance is far beyond the effective near-field coupling length in a normal environment, as is shown in Fig. 17(f). Recently, with the aid of high-$k$ modes excited in HMMs, the miniaturized and low lasing threshold laser has also been proposed [95, 264].

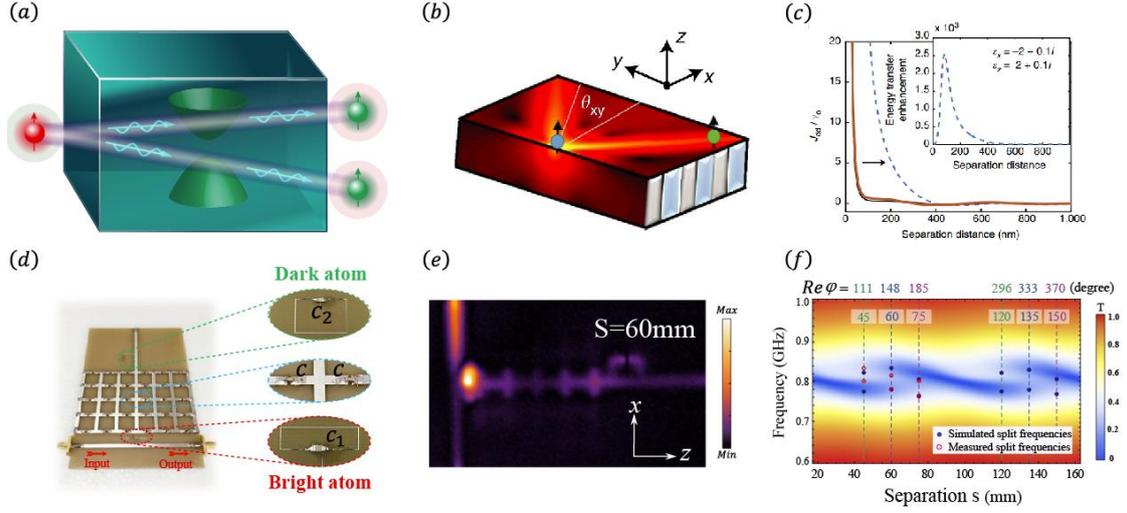

FIG. 17. (a) Schematic of the HMM assisted energy transfer between the bright and dark atoms. (b) Long-range dipole-dipole interaction realized by the HMS. (c) The cooperative Lamb shift as a function of the separation distance along the resonance cone angle. Reproduced under the terms of a Creative Commons AttributionNonCommercial-NoDerivs 4.0 International License. Nat. Commun. 8, 14144 (2017). Copyright 2017, The Authors. (d) Circuit-based HMMs for realizing EIT between dark atom and bright atom. (e) Measured electric fields show the long-range interaction between bright atom and dark atom. (f) The dependence of HMM-mediated EIT on $s$. The red open (blue solid) circles are the measured (simulated) EIT split frequencies. Reproduced with permission from Opt. Express 26, 627 (2018). Copyright 2018, OSA Publishing.

### 5.3 High sensitivity sensors

HMMs can be utilized to design various highly sensitive sensors such as the refractive index sensors by exciting the high-$k$ modes [265-271]. The highly localized bulk plasmon polariton in HMMs with large momentum is sensitive to the change of refractive index and has high sensitivity in biosensing applications. Two important HMM-based sensors in visible and near-infrared range have been demonstrated based on the metal/dielectric multilayered structure and metal nanowire array [265, 267]. For example, the miniaturized biosensor can be fabricated based on the gold/$Al_2O_3$ multilayered structure, as is shown in Fig. 18(a). The thickness of gold and $Al_2O_3$ layers are 16 nm and 30 nm, respectively [267]. Based on EMT, the effective anisotropic permittivity tensor is shown in Fig. 18 (b). When $520 \leq \lambda \leq 800$ nm, the sign of $\varepsilon_\perp$ and $\varepsilon_{//}$ are opposite and thereby the multilayered structure shows the hyperbolic dispersion. The high-$k$ modes in HMM is non-radiative and it cannot be directly excited because of the momentum mismatch. Therefore, a grating or prism needs to be placed above the structure to provide additional transverse wavevectors. For the TM wave incident on the structure, the reflectance spectrum at different incident angle are shown in Fig. 18(c). There are six prominent reflectance dips in the spectrum, which correspond to the six guided modes in the HMMs. Specially, two guided modes below 500 nm are the SPP modes instead of the high-$k$ modes in the HMMs. The physical mechanism of this refractive index sensors in HMMs is simple. Once the refractive index of the surrounding medium changes, the coupling condition will changes. This change can be directly determined by the shift of the resonance wavelength. By injecting different weight percentage concentrations of glycerol in distilled (DI) water, the reflectance spectrum of the HMM sensor is shown in Fig. 18(d). Figures 18(e) and 18(f) are the enlarged graph of Fig. 18(d) for the high-$k$ modes in HMM and SPP modes, respectively. By comparing Fig. 18(e) with 18(f), we can clearly see that high-$k$ mode in HMM has a significant shift of resonance wavelength, in which the HMM sensor achieved bulk refractive index sensitivity of 30000 nm/RIU (per refractive index unit). The record figure of merit (*FOM*) can determine the sensitivity of the sensor [267]:

$$Fom = (\Delta\lambda / \Delta n)(1 / \Delta\omega), \qquad (25)$$

where $\Delta\lambda$, $\Delta\omega$ and $\Delta n$ denote the shift of wavelength, and width of the reflectance dip at half-maximum and refractive index change, respectively. The greater the value of *FOM* is, the higher the sensitivity of the sensor is. The FOM in Figs. 18(e) and 18(f) are 590 and 108, respectively. So the high-$k$ mode in HMM is more sensitive to the extremely small refractive index change than the SPP [267]. In addition, the grating coupled-HMM biosensor can have higher angular sensitivity (with a maximum of 7000° per RIU) than the traditional sensors, which can be used to realize the angular detection of low molecular weight biomolecules [268]. Recently, the radiative refractive index sensors are proposed by using the phase singularity feature associated with the Brewster mode in HMMs [153,155].

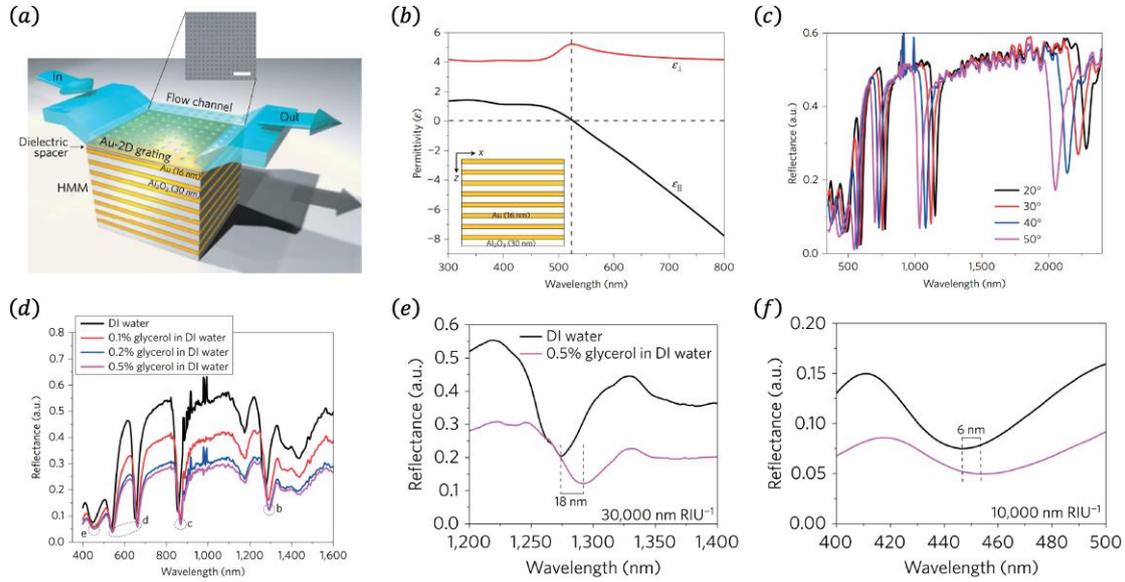

FIG. 18. (a) Schematic of the metal grating coupled HMM sensor with a fluid flow channel. (b) The effective permittivity of the metal/dielectric multilayer structure. (c) The reflectance of (a) at different incident angles. (d) The reflectance of (a) by injecting different weight percentage concentrations of glycerol in DI water. (e) Enlarged graph of (d) from 1200-1400 nm. (f) Similar to (e), but the spectrum from 400-500 nm. Reproduced with permission from Nat. Mater. 15, 621 (2016). Copyright 2016, Springer Nature.

## 6. Conclusion

HMMs have many fascinating properties, such as the enhanced spontaneous radiation, all-angle negative refraction, and abnormal scattering phenomena. Recently they have been widely studied based on the subwavelength metal/dielectric multilayered structure, metal nanowires array, the circuit system, the HMS, and even the natural hyperbolic media. The interesting properties of HMM are directly associated with the special hyperbolic dispersion. By tuning the topological transition of IFC, the interaction between light and matter will significantly be modified. In this tutorial, we mainly discuss three effective ways of manipulating the topological transition of IFCs such as changing the frequency, tuning the imaginary part of permittivity or permeability, and applying an external field. Specially, at the transition points of two kinds of topological transition, a very flat elliptic IFC and a linear crossing IFC are analyzed, respectively. In addition, the idea of dispersion manipulation of HMM has also be extended to the composite structure with HMMs. For example, the dispersionless bandgap can be realized in a 1D PHC based on the PVC condition. The dispersion manipulation in HMMs can be utilized in many applications such as the hyperlens breaking the diffraction limit, the long-range energy transfer beyond the near-field coupling limit, and the high sensitivity sensors. Because of the unusual properties of HMMs, it is expected that more and more optical devices based on HMMs would be realized in the future.

**ACKNOWLEDGMENTS**

This work is supported by the National Key R&D Program of China (Grant No. 2016YFA0301101), by the National Natural Science Foundation of China (NSFC) (Grants No. 11774261, No. 11474220, and No. 61621001), by the Natural Science Foundation of Shanghai (Grant No. 17ZR1443800), by the Shanghai Science and Technology Committee (Grant No. 18JC1410900) and by China Postdoctoral Science Foundation (Grant No. 2019TQ0232).